\documentclass[12pt]{article}
\usepackage{graphicx}  
\usepackage{setspace}
\usepackage{placeins}
\usepackage{floatpag}
\usepackage[utf8]{inputenc}
\usepackage{changepage}
\usepackage{nameref} 

\usepackage{amsmath}
\usepackage{bm}
\newcommand{\He}{\overline{H}}

\newcommand{\hrho}{\hat\rho_{_Y,_R}} 
\newcommand{\E}{{\mbox{E}}}

\usepackage{array}
\usepackage{booktabs}
\newcolumntype{L}[1]{>{\raggedright\let\newline\\\arraybackslash\hspace{0pt}}m{#1}}
\newcolumntype{C}[1]{>{\centering\let\newline\\\arraybackslash\hspace{0pt}}m{#1}}
\newcolumntype{R}[1]{>{\raggedleft\let\newline\\\arraybackslash\hspace{0pt}}m{#1}}

\usepackage{times}
\usepackage[T1]{fontenc}
\usepackage{newtxtext,newtxmath}
\usepackage[scaled=0.8]{beramono}

\usepackage{hyperref}
\hypersetup{colorlinks = true,
            urlcolor = crimson,
            linkcolor = crimson,
            citecolor = crimson}

\usepackage[labelfont=bf,figurename=Fig.]{caption}
\DeclareCaptionLabelSeparator{bar}{ | }
\usepackage[super,numbers,sort&compress]{natbib} 
\usepackage{chapterbib}

\usepackage{xcolor}
\definecolor{ax}{HTML}{cf7a30}
\definecolor{fb}{HTML}{4891dc}
\definecolor{hp}{HTML}{69913b}
\definecolor{crimson}{RGB}{156,0,0}

\title{\vspace{-0.5in}\textbf{Unrepresentative Big Surveys\\Significantly Overestimate US Vaccine Uptake}\bigskip}

\author
{Valerie C. Bradley$^{1\ast}$, Shiro Kuriwaki$^{2\ast}$, Michael Isakov$^{3}$,\\ Dino Sejdinovic$^{1}$, Xiao-Li Meng$^{4}$, Seth Flaxman$^{5,\dagger}$\\
\\
\normalsize{$^{1}$Department of Statistics, University of Oxford, Oxford, UK}\\
\normalsize{$^{2}$Department of Political Science, Stanford University, Stanford, CA, USA}\\
\normalsize{$^{3}$Harvard College, Harvard University, Cambridge, MA, USA}\\
\normalsize{$^{4}$Department of Statistics, Harvard University, Cambridge, MA, USA}\\
\normalsize{$^{5}$Department of Computer Science, University of Oxford, Oxford, UK}\\
\normalsize{$^\ast$These authors contributed equally to this work.}\\
\normalsize{$^{\dagger}$Correspondence to: \href{mailto:seth.flaxman@cs.ox.ac.uk}{\texttt{seth.flaxman@cs.ox.ac.uk}}.}
}
\date{\normalsize October  2021}

\begin{document} 

\baselineskip24pt

\maketitle 
\thispagestyle{empty}

\begin{abstract}
    Surveys are a crucial tool for understanding public opinion and behavior, and their accuracy depends on maintaining statistical representativeness of their target populations by minimizing biases from all sources. Increasing data size shrinks confidence intervals but magnifies the impact of survey bias -- an instance of the Big Data Paradox\cite{Meng2018}. Here we demonstrate this paradox in estimates of first-dose COVID-19 vaccine uptake in US adults: Delphi-Facebook\cite{Barkay2020,Kreuter2020} (about 250,000 responses per week) and Census Household Pulse\cite{censushp} (about 75,000 per week). By May 2021, Delphi-Facebook overestimated uptake by 17 percentage points and Census Household Pulse by 14, compared to a benchmark from the Centers for Disease Control and Prevention (CDC). Moreover, their large data sizes led to minuscule margins of error on the incorrect estimates. In contrast, an Axios-Ipsos online panel\cite{axiosipsos} with about 1,000 responses following survey research best practices\cite{aaporbest} provided reliable estimates and uncertainty. We decompose observed error using a recent analytic framework\cite{Meng2018} to explain the inaccuracy in the three surveys. We then analyze the implications for vaccine hesitancy and willingness. We show how a survey of 250,000 respondents can produce an estimate of the population mean that is no more accurate than an estimate from a simple random sample of size 10. Our central message is that data quality matters far more than data quantity, and compensating the former with the latter is a mathematically provable losing proposition.
\end{abstract}

\newpage

\addtocontents{toc}{\protect\setcounter{tocdepth}{0}} 

Governments, businesses, and researchers rely on survey data to inform the provision of government services\cite{hastak2001role}, steer business strategy, and guide response to the COVID-19 pandemic\cite{MMWRmay,arrieta2021}. With the ever-increasing volume and accessibility of online surveys and organically-collected data, the line between traditional survey research and Big Data is becoming increasingly blurred\cite{japec2015aapor}.
Large datasets enable analysis of fine-grained subgroups, which are in high-demand for designing targeted policy interventions\cite{reinhart2021vaxintro}. However, counter to common intuition\cite{mayer2013big}, larger sample sizes alone do not ensure lower error. Instead, small biases are \textit{compounded} as sample size increases\cite{Meng2018}.

We see initial evidence of this in the discrepancies in estimates of first-dose COVID-19 vaccine uptake, willingness, and hesitancy from three online surveys in the US. Two of them --- Delphi-Facebook's COVID-19 symptom tracker\cite{Barkay2020,Kreuter2020} ($n \approx$ 250,000 per week and with over 4.5 million responses from January to May 2021) and the Census Bureau's Household Pulse survey\cite{censushp} ($n \approx$ 75,000 per survey wave and with over 600,000 responses from January to May 2021) --- have large enough sample sizes to
render standard uncertainty intervals negligible, yet report significantly different estimates of vaccination behavior with nearly identically-worded questions (Table \ref{table:methodologies}).  For example, Delphi-Facebook's state-level estimates for willingness to receive a vaccine from the end of March 2021 are 8.5 percentage points lower on average than those from the Census Household Pulse (Extended Data Fig.~\ref{fig:hesitancy-state-march}A), with differences as large as 16 percentage points.

The US Centers for Disease Control and Prevention (CDC) compiles and reports vaccine uptake statistics from state and local offices\cite{cdc}.  These figures serve as a rare external benchmark, permitting us to compare survey estimates of vaccine uptake to those from the CDC. The CDC has noted the discrepancies between their own reported vaccine uptake and that of the Census Household Pulse\cite{cdcNguyen2021, cdcSantibanez}, and we find even larger discrepancies between the CDC and Delphi-Facebook data (Fig~\ref{fig:main}a). 
In contrast, the Axios-Ipsos' Coronavirus Tracker \cite{axiosipsos} ($n \approx$ 1,000 responses per wave, and over 10,000 responses from January to May 2021) tracks the CDC benchmark well. 
None of these surveys use the CDC benchmark to adjust or assess their estimates of vaccine uptake, thus by examining patterns in these discrepancies, we can infer each survey's accuracy and \textit{statistical representativeness}, a nuanced concept that is critical for the reliability of survey findings\cite{kruskal1979representativeI,kruskal1979representativeII,kruskal1979representativeIII,kruskal1980representativeIV}.


\section*{The Big Data Paradox in vaccine uptake}\label{sec:paradox}

We focus on the Delphi-Facebook and Census Household Pulse surveys because their large sample sizes (each greater than 10,000 respondents\cite{aaporlarge})  present the opportunity to examine the Big Data Paradox\cite{Meng2018} in surveys. The Census Household Pulse is an experimental product designed to rapidly measure pandemic-related behavior. Delphi-Facebook has stated that the intent of their survey is to make comparisons over space, time, and subgroups and that point estimates should be interpreted with caution\cite{Kreuter2020}. However, despite these intentions, Delphi-Facebook has reported point estimates of vaccine uptake in its own publications\cite{cmu2021topline,reinhart2021vaxintro}.

Delphi-Facebook and Census Household Pulse surveys persistently overestimate vaccine uptake relative to the CDC's benchmark (Fig.~\ref{fig:main}a). Despite being the smallest survey by an order of magnitude, Axios-Ipsos' estimates track well with the CDC rates (see Fig~\ref{fig:main}a), and their 95\% confidence intervals contain the benchmark estimate from the CDC in $10$ out of $11$ surveys (an empirical coverage probability of $91 \%$).

One might hope that estimates of changes in first-dose vaccine uptake are correct, even if each snapshot is biased. Unfortunately, errors have increased over time, from just a few percentage points in January 2021 to 4.2 percentage points (Axios-Ipsos), 14 percentage points (Census Household Pulse), and 17 percentage points (Delphi-Facebook) by mid-May 2021 (Fig.~\ref{fig:main}b).  For context, for a state near the herd immunity threshold (70-80\% based on recent estimates \cite{haas2021impact}), a discrepancy of 10 percentage points in vaccination rates could be the difference between containment and uncontrolled exponential growth in new SARS-CoV-2 infections.

Conventional statistical formulas for uncertainty further mislead when applied to biased big surveys because as sample size increases, bias (rather than variance) dominates estimator error. Fig.~\ref{fig:main}a shows 95\% confidence intervals for vaccine uptake based on each survey's reported sampling standard errors and weighting design effects \cite{Kish1965}. Axios-Ipsos has the widest confidence intervals, but also the smallest design effects (1.1-1.2), suggesting that its accuracy is driven more by minimizing bias in data collection rather than post-survey adjustment. Census Household Pulse's 95\% confidence intervals are widened by large design effects (4.4-4.8) but they are still too narrow to include the true rate of vaccine uptake in almost all survey waves. The confidence intervals for Delphi-Facebook are vanishingly small, driven by large sample size and moderate design effects (1.4-1.5), and give us essentially zero chance of being even close to the truth.  

One benefit of such large surveys might be to compare estimates of spatial and demographic subgroups\cite{ihme2021map,king2021time,cdc2021hesitant}. However, in March of 2021, Delphi-Facebook and Census Household Pulse over-estimated CDC state-level vaccine uptake by 16 and 9 percentage points, respectively (Extended Data Fig.~\ref{fig:hesitancy-state-march}G-H), and by equal or larger amounts by May 2021 (Extended Data Fig.~\ref{fig:hesitancy-state-may}G-H). Relative estimates were no better than absolute estimates in March of 2021: there is barely any agreement in a survey's estimated state-level rankings with the CDC (a Kendall rank correlation of 0.31 for Delphi-Facebook in Extended Data Fig.~\ref{fig:hesitancy-state-march}I, 0.26 for Census Household Pulse in Extended Data Fig.~\ref{fig:hesitancy-state-march}J) but they improved in May of 2021 (correlations 0.78 and 0.74 in Extended Data Fig.~\ref{fig:hesitancy-state-may}I-J). Among 18-64 year-olds, both Delphi-Facebook and Census Household Pulse overestimate uptake, with errors increasing over time (Extended Data Fig.~\ref{fig:ddc_by-age}).

These examples illustrate a mathematical fact. That is, 
when biased samples are large, they are doubly misleading: they produce confidence intervals with incorrect centers and substantially underestimated widths. This is the Big Data Paradox\cite{Meng2018}: \textit{the larger the data size, the surer we fool ourselves} when we fail to account for bias in data collection.

\section*{A framework for quantifying data quality}\label{sec:ddc}

While it is well-understood that traditional confidence intervals capture only survey sampling errors\cite{groves2011survey} (and not total error), tools for quantifying nonsampling errors separately from sampling errors are difficult to apply and rarely used in practice.
A recently formulated  statistical framework \cite{Meng2018} permits us to exactly decompose total error of a survey estimate into three components:
\begin{equation}
   \mbox{\textbf{Total Error}} = {\textbf{Data Quality Defect}} \times{\textbf{Data Scarcity}} \times {\textbf{Inherent Problem Difficulty}}. \label{eq:meng_identity}
\end{equation}
This framework has been applied to COVID-19 case counts \cite{Dempsey2020} and election forecasting \cite{isakov2020towards}. Its full application requires ground-truth benchmarks or their estimates from independent sources\cite{Meng2018}. 

Specifically, {\bf Total Error} is the difference between the observed sample mean $\bar Y_n$ as an estimator of the ground truth, the population mean $\bar Y_{_N}$. The {\bf Data Quality Defect} is measured using
$\hat\rho_{_Y, _R}$,  called  \textit{data defect correlation} (\textit{ddc})\cite{Meng2018}, which quantifies total bias (from any source), measured by the correlation between the event that an individual's response is recorded and its value, $Y$. The impact of data quantity is captured by {\bf Data Scarcity}, which is a function of the sample size $n$ and the population size $N$, measured as $\sqrt{\frac{N -n}{n}}$, and hence what matters for error is the relative sample size, i.e.,~how close $n$ is to $N$, rather the absolute sample size $n$. The third factor is 
{\bf Inherent Problem Difficulty}, which measures the population heterogeneity (via standard deviation $\sigma_{_Y}$ of $Y$), because the more heterogeneous a population is, the harder it is to estimate its average well. Mathematically, Equation \eqref{eq:meng_identity} is given by $\bar{Y}_n - \bar{Y}_{_N} = \hat\rho_{_Y, _R} \times \sqrt{\frac{N -n}{n}} \times \sigma_{_Y}$. Incidentally, this expression was inspired by the Hartley-Ross inequality for biases in ratio estimators published in \textit{Nature} in 1954\cite{hartley1954unbiased}. More details on the decomposition are provided in the Methods (``\nameref{sec:methods-ddc}''), where we also present a generalization for weighted estimators.

\section*{Decomposing error in COVID surveys}\label{sec:emperical}

\begin{figure}[tp]
    \centering
\begin{adjustwidth}{-3em}{-3em}
    \includegraphics[width=18cm]{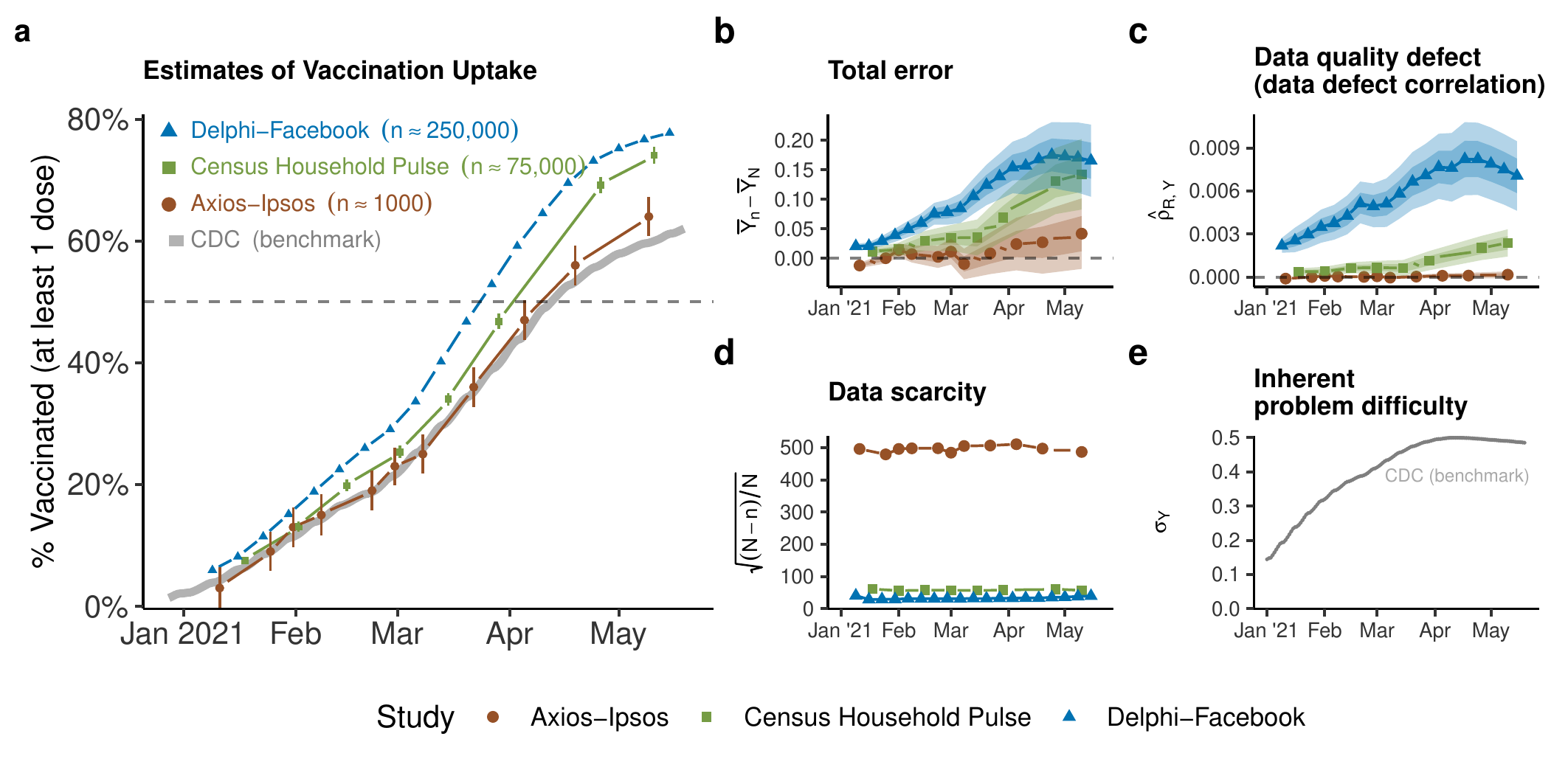}
\end{adjustwidth}
    \caption{\textbf{Errors in estimates of vaccine uptake}.
    \textbf{a.} Estimates of vaccine uptake for US adults in 2021 compared to CDC benchmark data, plotted by end date of each survey wave. Points indicate each study's weighted estimate of first-dose vaccine uptake, and intervals are 95\% CIs using reported standard errors and design effects. Delphi-Facebook has $n=4,525,633$ across 19 waves, Census Household Pulse has $n=606,615$ across 8 waves, and Axios-Ipsos has $n=11,421$ across 11 waves. Delphi-Facebook's CIs are too small to be visible.   
    \textbf{b.} Total error \(\bar{Y}_{n} - \bar{Y}_{_N}\), \textbf{c.} data defect correlation $\hat\rho_{_Y, _R}$,
    \textbf{d.} data scarcity \(\sqrt{(N-n)/n}\), \textbf{e.} inherent problem difficulty \(\sigma_{_Y}\).
     Shaded bands represent scenarios of +/-5\% (darker) and +/-10\% (lighter) error in the CDC benchmark relative to reported values (points). \textbf{b} - \textbf{e} comprise the decomposition in Equation \eqref{eq:meng_identity}.
    \label{fig:main}
    }
     \thisfloatpagestyle{empty}
\end{figure}

While the data defect correlation \textit{ddc} is not directly observed, COVID-19 surveys present a rare case in which it can be deduced because all other terms in Equation \eqref{eq:meng_identity} are known (see ``\nameref{sec:methods-ddc}'' in the Methods for an in-depth explanation).
 We apply this framework to the aggregate error shown in Fig.~\ref{fig:main}b, and the resulting components of error from the right-hand side of Equation~\ref{eq:meng_identity} are shown in Fig.~\ref{fig:main}c-e.

We use the CDC's report of the cumulative count of first doses administered to US adults as the benchmark\cite{cdc,MMWRmay}, $\bar{Y}_{_N}$.
This benchmark may suffer from administrative delays and slippage in how the CDC centralizes information from states \cite{tiu2021,groen2012,tu1993aids,ft2021benchmark}. As a sensitivity analysis to check the robustness of our findings to further misreporting, we present our results with sensitivity intervals under the assumption that CDC's reported numbers suffer from $\pm 5\%$ and $\pm10\%$ error. These scenarios were chosen based on analysis of the magnitude by which the CDC's initial estimate for vaccine uptake by a particular day increases as the CDC receives delayed reports of vaccinations that occurred on that day (Extended Data Fig.~\ref{fig:benchmark-change} and Supplementary Information \ref{sec:SM-benchmark}). That said, these scenarios may not capture latent systemic issues affecting CDC vaccination reporting.

The \textbf{Total Error} of each survey's estimate of vaccine uptake (Fig.~\ref{fig:main}b) increases over time for all studies, most markedly for Delphi-Facebook. 
The \textbf{Data Quality Defect}, measured by the \textit{ddc}, also increases over time for Census Household Pulse and for Delphi-Facebook (Fig.~\ref{fig:main}c). The \textit{ddc} for Axios-Ipsos is much smaller and steady over time, consistent with what one would expect from a representative sample. The \textbf{Data Scarcity} ($\sqrt{\frac{N-n}{n}}$) for each survey is roughly constant across time (Fig.~\ref{fig:main}d). \textbf{Inherent Problem Difficulty} is a population quantity common to all three surveys which peaks when the benchmark vaccination rate approaches 50\% in April 2021 (Fig.~\ref{fig:main}e).
Therefore, the decomposition suggests that the increasing error in estimates of vaccine uptake in Delphi-Facebook and Census Household Pulse is primarily driven by increasing \textit{ddc}, which captures the overall impact of the bias in coverage, selection, and response. 

Equation \eqref{eq:meng_identity} also yields a formula for the bias-adjusted effective sample size $n_\text{eff}$, which is the size of a simple random sample that we would expect to exhibit the same level of Mean Square Error (MSE) as what was actually observed in a given study with a given \textit{ddc}. Unlike the classical effective sample size \cite{Kish1965}, this quantity captures the impact of bias as well as that of an increase in variance from weighting and sampling. Details for this calculation are in Methods (\nameref{sec:methods-weighted}). 

\begin{figure}[tp]
    \centering
    \includegraphics[width = 9cm]{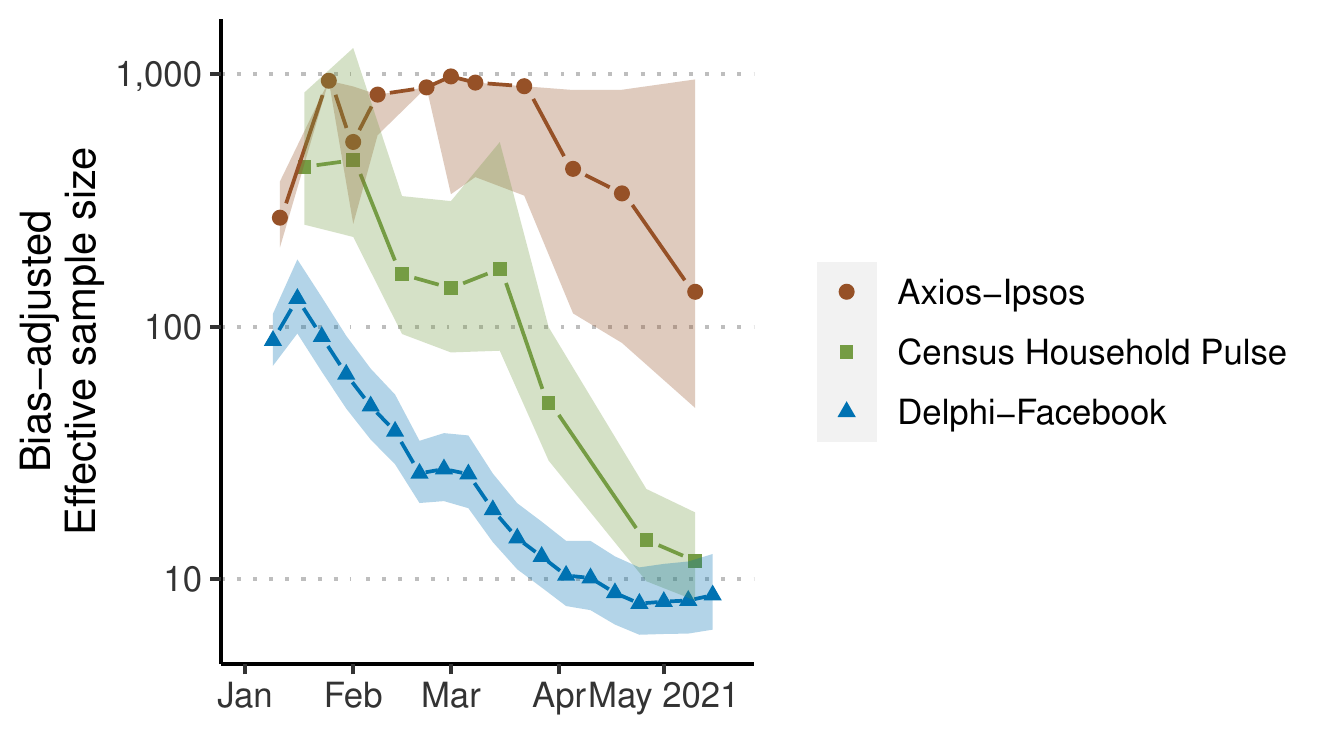}
    \caption{\textbf{Bias-adjusted effective sample size.} The bias-adjusted effective sample size of an estimate (different from the classic Kish effective sample size) is the size of a simple random sample which would have the same Mean Square Error as the observed estimate. 
    Effective sample sizes are shown on the $\log_{10}$ scale.
    The original sample size was $n = $ 4,525,633 across 19 waves for Delphi-Facebook, $n = $ 606,615 across 8 waves for Census Household Pulse, $n = $ 11,421 across 11 waves for Axios-Ipsos.
    Shaded bands represent scenarios of +/-5\% error in the CDC benchmark relative to point estimates based on actual reported values.}
    \label{fig:n-eff}
\end{figure}

For estimating the US vaccination rate, Delphi-Facebook has a bias-adjusted effective sample size of less than 10 in April 2021, a 99.99\% reduction from the raw average weekly sample size of 250,000 (Fig.~\ref{fig:n-eff}). The Census Household Pulse also suffers from over 99\% reductions in effective sample size by May 2021. A simple random sample would have controlled estimation errors by controlling \textit{ddc}. However, once this control is lost, small increases in \textit{ddc} beyond what is expected in simple random samples can result in drastic reductions of effective sample sizes for large populations \cite{Meng2018}.

\FloatBarrier

\section*{Comparing study designs}

Understanding \textit{why} bias occurs in some surveys but not others requires an understanding of the sampling strategy, modes, questionnaire, and weighting scheme of each survey. Table~\ref{table:methodologies} compares the design of each survey (more details in the ``\nameref{sec:SM-data-background}'' in Methods and Extended Data Table \ref{table:methodologies-ED}).

\begin{table}
\begin{adjustwidth}{-1cm}{-1cm}
\footnotesize
    \centering
    \begin{tabular}{L{3cm}  L{4.6cm}  L{4.6cm}  L{4.6cm} }
    \toprule
        & \textcolor{ax}{\textbf{Axios-Ipsos}}   & \textcolor{hp}{\textbf{Census Household Pulse}} & \textcolor{fb}{\textbf{Delphi-Facebook}}  \\ \midrule
        \textbf{Recruitment mode} & Address-based mail sample to Ipsos KnowledgePanel & SMS and email & Facebook Newsfeed \\ \midrule
        \textbf{Interview mode} & Online &  Online & Online \\ \midrule
        \textbf{Average size} & 1,000/wave & 75,000/wave & 250,000/week \\ \midrule
        \textbf{Sampling frame} & Ipsos KnowledgePanel; internet/tablets provided to $\sim$5\% of panelists who lack home internet  & Census Bureau’s Master Address File (individuals for whom email / phone contact information is available) & Facebook active users  \\ \midrule
        \textbf{Vaccine uptake question} & ``Do you personally know anyone who has already received the COVID-19 vaccine?'' & ``Have you received a COVID-19 vaccine?'' & ``Have you had a COVID-19 vaccination?'' \\ \hline
        \textbf{Vaccine uptake definition} & ``Yes, I have received the vaccine''  &  ``Yes'' &  ``Yes''  \\ \hline
        \textbf{Other response options} & ``Yes, a member of my immediate family,'' ``Yes, someone else,'' 'No' &  ``No'' &  ``No,'' ``I don't know''  \\ \hline
        \textbf{Weighting variables} & %
        Gender by age, race, education, Census region, metropolitan status, household income, partisanship. &  %
        Education by age by sex by state, race/ethnicity by age by sex by state, household size &  %
        Stage 1: age, gender ``other attributes which we have found in the past to correlate with survey outcomes'' to FAUB; Stage 2: state by age by gender \\
        \bottomrule
    \end{tabular}
    \caption{\textbf{Comparison of survey designs.} Comparison of key design choices across Axios-Ipsos, Census Household Pulse, and Delphi-Facebook studies. All surveys target the US adult population. See Extended Data Table \ref{table:methodologies-ED} for additional comparisons.}
    \label{table:methodologies}
\end{adjustwidth}
\end{table}

All three surveys are conducted online and target the US adult population, but vary in respondent recruitment methods\cite{pew:noprobability}. The Delphi-Facebook survey recruits respondents from active Facebook users (the Facebook Active User Base, or FAUB) using daily unequal-probability stratified random sampling\cite{Barkay2020}. The Census Bureau uses a systematic random sample to select households from the subset of the Census' Master Address File (MAF) for which they have obtained either cell phone or email contact information (approximately 81\% of all households on the MAF)\cite{censushp}.  

In comparison, Axios-Ipsos relies on inverse response propensity sampling from Ipsos' online KnowledgePanel. Ipsos recruits panelists using an address-based probabilistic sample from USPS's Delivery Sequence File (DSF)\cite{axiosipsos}. The DSF is similar to the Census' MAF. Unlike the Census Household Pulse, potential respondents are not limited to the subset for whom email and phone contact information is available. Furthermore, Ipsos provides internet access and tablets to recruited panelists who lack home internet access. In 2021, this ``offline'' group typically comprises 1\% of the final survey (Extended Data Table \ref{tab:ax-offline}).

All three surveys weight on age and gender, i.e., assign larger weights to respondents of underrepresented age by gender subgroups and smaller weights to those of overrepresented subgroups\cite{Barkay2020,censushp,axiosipsos} (Table~\ref{table:methodologies}). Axios-Ipsos and Census Household Pulse also weight on education and race/ethnicity. Axios-Ipsos additionally weights to the composition of political partisanship measured with the ABC News/Washington Post poll in 6 of the 11 waves we study. Education, a known correlate of propensity to respond to surveys \cite{Kennedy2018} and social media use \cite{pew:socialmedia}, are notably absent from Delphi-Facebook's weighting scheme, as is race/ethnicity. As noted before, none of the surveys use the CDC benchmark to adjust or assess estimates of vaccine uptake.

\section*{Explanations for error}\label{sec:why-error}

Table \ref{tab:edu-nonrepresentative} illustrates some consequences of these design choices. Axios-Ipsos samples mimic the actual breakdown of education attainment among US adults even before weighting, while those of Census Household Pulse and Delphi-Facebook do not. After weighting, Axios-Ipsos and Census Household Pulse match the population benchmark, by design. Delphi-Facebook does not explicitly weight on education, and hence the education bias persists in their weighted estimates: those without a college degree are underrepresented by nearly 20 percentage points. The story is similar for race/ethnicity. Delphi-Facebook's weighting scheme does not adjust for race/ethnicity, and hence their weighted sample still over-represents White adults by 8 percentage points, and under-represents Black and Asian proportions by around 50 percent of their size in the population (Table~\ref{tab:edu-nonrepresentative}). 

\begin{table}
\footnotesize
\centering

\begin{tabular}{R{2.2cm} cc cc cc c ccc}
\toprule
 & \multicolumn{7}{c}{Composition of US Adults}
 & \multicolumn{3}{c}{Survey Estimates}\\
 \cmidrule(lr){2-8} \cmidrule(lr){9-11}
    &  \multicolumn{2}{c}{Axios-Ipsos} &
    \multicolumn{2}{c}{Household Pulse} & \multicolumn{2}{c}{Delphi-Facebook} & 
    ACS & 
    \multicolumn{3}{c}{Household Pulse}\\
    \cmidrule(lr){2-3} \cmidrule(lr){4-5}  \cmidrule(lr){6-7}  \cmidrule(lr){9-11}
    & Raw & Weighted &  Raw & Weighted & Raw & Weighted & Benchmark  & Vax & Will & Hes\\
    \textbf{Education} & \\\midrule
    High School    & 35\% & 39\% & 14\% & 39\% & 19\% & 21\% & 39\%
 & 39\% & 40\%&  21\%\\
    Some College   & 29 & 30 & 32 & 30 & 36 & 36 & 30
 & 44 & 38 &   18\\
    4-Year College & 19 & 17 & 29 & 17 & 25 & 25 & 19
 & 54 & 36  & 10 \\
    Post-Graduate  & 17 & 14 & 26 & 13 & 20 & 18 & 11
 &  67 & 26 & 7 \\
    \textbf{Race/Ethnicity} &\\\midrule
    White    & 71\% & 63\% & 75\% & 62\% & 74\% & 68\% & 60\%
  & 50\% & 33\% & 17\% \\
    Black    & 10 & 12 & 7 & 11 & 6 & 6 & 12
 & 42 & 39 & 19\\
    Hispanic & 11 & 16 & 10 & 17 & 11 & 16 & 16
 & 38 & 48 & 14\\
    Asian    &  &  & 5 & 5 & 2 & 3 & 6
 & 51 & 43 & 5\\\bottomrule
\end{tabular}
\caption{ \textbf{Composition of survey respondents by educational attainment and race/ethnicity.}  Axios-Ipsos: wave ending March 22, 2021, $n$  = 995. Census Household Pulse: wave ending March 29, 2021, $n$ = 76,068.  Delphi-Facebook: wave ending March 27, 2021, $n$ = 181,949. Benchmark uses the 2019 US Census American Community Survey (ACS), composed of roughly 3 million responses. Right-most column shows estimates of vaccine uptake (Vax), willingness (Will) and hesitancy (Hes) from the Census Household Pulse of the same wave. }
\label{tab:edu-nonrepresentative}
\end{table}

The overrepresentation of White adults and people with college degrees explains part of the error of Delphi-Facebook. The racial groups that Delphi-Facebook under-represents tend to be more willing and less vaccinated in the samples (Table~\ref{tab:edu-nonrepresentative}). In other words, re-weighting the Delphi-Facebook survey to upweight racial minorities will bring willingness estimates closer to Household Pulse and the vaccination rate closer to CDC. The three surveys also report that people \textit{without} a 4-year college degree are less likely to have been vaccinated compared to those \textit{with} a degree (Table \ref{tab:edu-nonrepresentative} and Supplemental Information Table \ref{tab:outcome-by-demo}). If we assume that vaccination behaviors do not differ systematically between non-respondents and respondents \emph{within} each demographic category, under-representation of less-vaccinated groups would contribute to the bias found here.  However, this alone cannot explain the discrepancies in all the outcomes. Census Household Pulse weights on both race and education\cite{censushp} and still over-estimates vaccine uptake by over ten points in late May of 2021 (Fig.~\ref{fig:main}b).

Delphi-Facebook and Census Household Pulse may be unrepresentative with respect to political partisanship, which has been found to be correlated with vaccine behavior \cite{gadarian2021partisanship} and with survey response \cite{pew2018optin}, and thus may contribute to observed bias.
However, neither Delphi-Facebook nor Census Household Pulse collects partisanship of respondents; Census agencies are prohibited from asking about political preference.
Moreover, no unequivocal population benchmark for partisanship exists.

Rurality may also contribute to the errors because it  correlates with vaccine status \cite{MMWRmay} and home internet access\cite{ryan2016}. Neither the Census Household Pulse nor Delphi-Facebook weights on sub-state geography, which may mean that adults in more rural areas are less likely to be vaccinated and also underrepresented in the surveys, leading to overestimation of vaccine uptake.

Axios-Ipsos weights to metropolitan status and also recruits a fraction of its panelists from an  ``offline'' population of individuals without Internet access. 
We find that \emph{dropping} these offline respondents ($n = 21$, or 1 percent of the sample) in their March 22, 2021 wave \emph{increases} Axios-Ipsos' overall estimate of the vaccination rate by 0.5 percentage points, thereby increasing the total error (Extended Data Table \ref{tab:ax-offline}). However,
this offline population is simply too small to explain the entirety of the difference in accuracy between Axios-Ipsos and either the Census Household Pulse (6 percentage points) or Delphi-Facebook (14 percentage points), in this time period. 

Careful recruitment of panelists is at least as important as weighting. 
Weighting on observed covariates alone cannot explain or correct the discrepancies we observe. For example, reweighting Axios-Ipsos March 22, 2021 wave using only Delphi-Facebook's weighting variables (age group and gender) increased the error in their vaccination estimates by 1 percentage point, but this estimate with Axios-Ipsos data is still more accurate than that from Delphi-Facebook during the same period (Extended Data Table \ref{tab:ax-offline}). The Axios-Ipsos estimate with Delphi-Facebook weighting overestimated vaccination by 2 percentage points, whereas Delphi-Facebook overestimated it by 11 percentage points.

The key implication is that there is no silver bullet: every small part of panel recruitment, sampling, and weighting matters for controlling the data quality measured as the correlation between an outcome and response, what we call the \emph{ddc}.  In multi-stage sampling, which includes for instance the selection of participants followed by non-response, bias in even a \emph{single} step can substantially impact the final result (see Methods ``\nameref{sec:multipopulation}'', Extended Data Table \ref{tab:census-ddc-stage}). A \textit{total quality control} approach -- inspired by the Total Survey Error framework\cite{biemer2003introduction} -- is a better strategy than trying to prioritize some components over others in order to improve data quality.  This emphasis is merely a reaffirmation of the best practice for survey research as advocated by the {American Association for Public Opinion Research}\cite{aaporbest}: ``The quality of a survey is best judged not by its size, scope, or prominence, but by how much attention is given to [preventing, measuring and] dealing with the many important problems that can arise.''\cite{scheuren2004survey}

\section*{Addressing common misperceptions}\label{sec:miss}

The three surveys discussed in this article demonstrate a seemingly paradoxical phenomenon -- the two larger surveys that we studied are far more statistically confident, yet also far more biased, than the smaller, more traditional Axios-Ipsos poll. These findings are paradoxical only when we fall into the trap of the long-held, but incorrect, intuition that estimation errors necessarily decrease in larger datasets \cite{mayer2013big}.

A limitation of our vaccine uptake analysis is that we only examine \textit{ddc} with respect to an outcome for which a benchmark is available: first-dose vaccine uptake. One might hope that surveys biased on vaccine uptake are not biased on other outcomes, for which there may not be benchmarks to expose their biases. However, the absence of evidence of bias for the remaining outcomes is not evidence of its absence. In fact, mathematically, 
when a survey is found to be biased with respect to one variable, it implies that the entire survey fails to be \textit{statistically representative}.
The theory of survey sampling relies on statistical representativeness for all variables achieved via probabilistic sampling\cite{sukhatme1954sampling}. 
Indeed, Neyman's original introduction of probabilistic sampling showed the limits of purposive sampling, which attempted to achieve overall representativeness via enforcing it only on a set of variables\cite{Neyman, kruskal1979representativeIII}. 

In other words, when a survey loses its overall statistical representativeness (e.g., through bias in coverage or nonresponse), which is difficult to repair (e.g., via weighting or modeling on observable characteristics) and almost impossible to verify\cite{groves2006nonresponse}, researchers who wish to use the survey for scientific studies must supply other reasons to justify the reliability of their survey estimates, such as evidence about the independence between the variable of interest and the factors that are responsible for the unrepresentativeness. Furthermore, scientific journals that wish to publish studies based on unrepresentative surveys\cite{kruskal1979representativeII}, especially those with large sizes such as Delphi-Facebook (biased with respect to vaccination status (Fig.~\ref{fig:main}), race and education (Table~\ref{tab:edu-nonrepresentative})), need to ask for reasonable effort from the authors to address the unrepresentativeness. A simple acknowledgment of the potential bias is insufficient for alerting about potentially seriously flawed findings, as we reveal in this article.

Some may argue that bias is a necessary trade-off for having data that is sufficiently large for conducting highly granular analysis, such as county-level estimation of vaccine hesitancy\cite{cdc2021hesitant}. 
While high-resolution inference is important, we warn that this is a double-edged argument. 
A highly biased estimate with a misleadingly small confidence interval can do more damage than having no estimate at all. 
We further note that bias is not limited to population point estimates, but also affects estimates of changes over time (contrary to published guidance\cite{Kreuter2020}) -- both Delphi-Facebook and Census Household Pulse significantly overestimate the \textit{slope} of vaccine uptake relative to that of the CDC benchmark (Fig.~\ref{fig:main}b).

The accuracy of our analysis also relies on the accuracy of the CDC's estimates of COVID vaccine uptake. However, if the selection bias in the CDC's benchmark is significant enough to alter our results, then that itself would be yet another example of the Big Data Paradox.

\section*{Discussion}

This is not the first time that the Big Data Paradox has reared its head: Google Trends predicted more than twice the number of influenza-like illnesses than the CDC in February 2013 \cite{Lazer2014}.  This analysis demonstrates that the Big Data Paradox applies not only to organically-collected Big Data, like Google Trends, but also to surveys. Delphi-Facebook is ``the largest public health survey ever conducted in the United States''\cite{salomon2021us}. The Census Household Pulse is conducted in collaboration between the US Census Bureau and eleven statistical government partners, all with enormous resources and survey expertise. Both studies take steps to mitigate selection bias, yet overestimate vaccine uptake by double digits. As we demonstrated, the impact of bias is magnified as relative sample size increases.

In contrast, Axios-Ipsos records only about 1,000 responses per wave, but makes additional efforts to prevent selection bias. Small surveys can be just as wrong as large surveys in expectation -- of the three other small-to-medium online surveys additionally analyzed, two also miss the CDC vaccination benchmark (Extended Data  Fig. \ref{fig:morepolls}). The overall lesson is that investing in data quality (particularly during collection, but also in analysis) minimizes error more efficiently than does increasing data quantity. Of course, a sample size of 1,000 may be too small (i.e.,~leading to unhelpfully large confidence intervals) for the kind of 50-state estimates given by big surveys. However, small area methods that borrow information across subgroups\cite{park2004} can perform better with better quality, albeit small, data, and it is an open question whether that approach would outperform the large, biased surveys.

There are approaches to correct for these biases in both probability and nonprobability samples alike. 
For COVID-19 surveys in particular, since June 2021, the AP-NORC multi-mode panel has weighted their COVID-19 related surveys to the CDC benchmark, so that the weighted \emph{ddc} for vaccine uptake is zero by design\cite{APNORC-JUNE}.
More generally, there is an extensive literature on approaches for making inferences from data collected from nonprobability samples\cite{wang2015forecasting,elliott2017inference,little2020measures}.
Other promising approaches include integrating surveys of varying quality\cite{wisniowski2020, yang2020doubly}, and leveraging the estimated \textit{ddc} in one outcome to correct bias in others under several scenarios (Supplemental Information \ref{sec:SM-scenarios}). 

While more needs to be done to fully examine the nuances of large surveys, organically collected administrative datasets, and social media data,  we hope this first comparative study of \emph{ddc} highlights the alarming implications of the \textit{Big Data Paradox} -- how large sample sizes magnify the impact of seemingly small defects in data collection, leading to overconfidence in incorrect inferences.

\newpage

\setcounter{section}{0}
\setcounter{subsection}{0}
\renewcommand{\thesection}{M\arabic{section}}

\begin{center}
{\LARGE \textbf{Methods}}
\end{center}

\section*{Calculation and interpretation of  \textit{ddc}}\label{sec:methods-ddc}

The mathematical expression for Equation \eqref{eq:meng_identity} is given  here for completeness: 
\begin{equation}
\bar{Y}_n - \bar{Y}_{_N} = \hat\rho_{_Y, _R} \times \sqrt{\frac{N -n}{n}} \times \sigma_{_Y}
\label{eq:meng_identity_math}
\end{equation}
The first factor  $\hat\rho_{_Y, _R}$ is  called the  \textit{data defect correlation} (\textit{ddc})\cite{Meng2018}. It is a measure of data quality represented by the correlation between the recording indicator $R$ ($R=1$ if
an answer is recorded and $R=0$ otherwise) and its value, $Y$. 
Given a benchmark, the \textit{ddc} $\hat\rho_{_Y, _R}$ can be calculated by substituting known quantities into Equation \eqref{eq:meng_identity_math}. 
In the case of a single survey wave of a COVID-19 survey, $n$ is the sample size of the survey wave, $N$ is the population size of US adults from US Census estimates \cite{censuspopest}, $\bar{Y}_n$ is the survey estimate of vaccine uptake, and  $\bar{Y}_{_N}$ is the estimate of vaccine uptake for the corresponding period taken from the CDC's report of the cumulative count of first doses administered to US adults \cite{cdc,MMWRmay}. We calculate $\sigma_{_Y} = \sqrt{\bar{Y}_{_N} \left(1 - \bar{Y}_{_N}\right)}$ because $Y$ is binary (but Equation \eqref{eq:meng_identity_math} is not restricted to binary $Y$).

We calculate $\hrho$ by using \textit{total} error $\bar{Y}_n - \bar{Y}_{_N}$, which captures not only selection bias but also any measurement bias (e.g.,~from question wording). However, with this calculation method, $\hrho$ lacks the direct interpretation as a correlation between $Y$ and $R$, and instead becomes a more general index of data quality directly related to classical design effects (see Methods section ``\nameref{sec:methods-neff}'').

It is important to point out that the increase  in \emph{ddc} does not necessarily imply that the response mechanisms for Delphi-Facebook and Census Household Pulse have changed over time. The correlation between a changing \textit{outcome} and a steady response mechanism could change over time, hence changing the value of \textit{ddc}. For example, as more individuals become vaccinated, and vaccination status is driven by individual behavior rather than eligibility, the correlation between vaccination status and propensity to respond could increase even if propensity to respond for a given individual is constant. This would lead to large values of \textit{ddc} over time, reflecting the \textit{increased impact} of the same response mechanism.  

\section*{Error decomposition with survey weights}\label{sec:methods-weighted}
The data quality framework given by Equations \eqref{eq:meng_identity} and \eqref{eq:meng_identity_math} is a special case of a more general framework for assessing the actual error of a weighted estimator $\bar Y_w = \sum_{i} w_iR_iY_i/\sum_{i} w_iR_i$, where $w_i$ is the survey weight assigned to individual $i$. It is shown in Meng\cite{Meng2018} that 
\begin{equation}\label{eq:weiden}
    \bar{Y}_\text{w} - \bar{Y}_{_N} = \hat\rho_{Y, R_{\text{w}}} \times \sqrt{\frac{N - n_\text{w}}{n_\text{w}}} \times \sigma_{_Y},
\end{equation}
where $\hat\rho_{Y, R_{\text{w}}}=\textrm{Corr}(Y, R_{\text{w}})$ is the finite population correlation between $Y_i$ and $R_{\text{w},i}=w_iR_i$ (over $i=1, \ldots, N$). The ``hat'' on $\rho$ reminds us that this correlation depends on the specific realization of $\{R_i, i=1,\ldots, N\}$.  The term $n_\text{w}$ is the classical ``effective sample size'' due to weighting \cite{Kish1965}, i.e., $n_\text{w} = n/(1+ CV_{\text{w}}^2)$, where $CV_\text{w}$ is the coefficient of variation of the weights for all individuals in the observed sample, that is,  the standard deviation of weights normalized by their mean. 
It is common for surveys to rescale their weights to have mean 1, in which case $CV^2_{w}$ is simply the sample variance of $W$.

When all weights are the same, Equation \eqref{eq:weiden} reduces to Equation \eqref{eq:meng_identity_math}.  In other words, the \textit{ddc} term  $\hat\rho_{Y, R_{\text{w}}}$ now also takes into account the impact of the weights as a means to combat the selection bias represented by the recording indicator $R$.  Intuitively, if $\hat\rho_{Y,R} = \textrm{Corr}(Y, R)$ is high (in magnitude), then some $Y_i$'s have a higher chance of entering our data set than others, thus leading to a sample average that is a biased estimator for the population average. Incorporating appropriate weights can reduce $\hat\rho_{_Y,_R}$ to $\hat\rho_{Y, R_{\text{w}}}$, with the aim to reduce the impact of the selection bias. However, this reduction alone may not be sufficient to improve the accuracy of $\bar Y_w$ because the use of weight necessarily reduces the sampling fraction $f=n/N$ to $f_w=n_w/N$ as well since $n_w <n$. Equation \eqref{eq:weiden} precisely describes this trade off, providing a formula to assess when the reduction of \textit{ddc} is significant to outweigh the reduction of the effective sample size.  

Measuring the correlation between $Y$ and $R$ is not a new idea in survey statistics (though note that \textit{ddc} is the population correlation between $Y$ and $R$, not the sample correlation), nor is the observation that as sample size increases, error is dominated by bias instead of variance \cite{bethlehem2002,meng2014}. The new insight is that \textit{ddc} is a general metric to index the \textit{lack of} representativeness of the data we observe, regardless of whether or not the sample is obtained via a probabilistic scheme, or weighted to mimic a probabilistic sample. As discussed in the the section on addressing common missperception,  any single \textit{ddc} deviating from what is expected under representative sampling (e.g., probabilistic sampling) is sufficient to establish the sample is not representative (but the converse is not true). Furthermore, the \textit{ddc} framework refutes the common belief that increasing sample size necessarily improves statistical estimation\cite{meng2014got, Meng2018}.

\section*{Bias-adjusted effective sample size}\label{sec:methods-neff}

By matching the mean-squared error of  $\bar Y_w$ with the variance of the sample average from simple random sampling, Meng\cite{Meng2018} derives the following formula for calculating a \textit{bias-adjusted effective sample size}, or $n_\text{eff}$:
\begin{align*}
    n_\text{eff} = \frac{n_\text{w}}{N-n_\text{w}} \times \frac{1} {E[\hat\rho_{Y,R_{\text{w}}}^2]}
\end{align*}
Given an estimator $\bar{Y}_\text{w}$ with expected total Mean Squared Error (MSE) $T$ due to data defect, sampling variability, and weighting, this quantity $n_\text{eff}$ represents the size of a simple random sample such that its mean $\bar Y_{_N}$, as an estimator for the same population mean $\bar Y_{_N}$, would have the identical MSE $T$.
The term $E[\hat\rho_{Y,R_{\text{w}}}^2]$ represents the amount of selection bias (square) expected on average from a particular recording mechanism $R$ and a chosen weighting scheme.

For each survey wave,  we use $\hat{\rho}_{Y,R_{\text{w}}}^2$ to approximate $E[\hat\rho_{Y,R_{\text{w}}}^2]$. This estimation is unbiased by design, since we use an estimator to estimate its expectation. Therefore, the only source of error is the sampling variation, which is typically negligible for large surveys, such as for Delphi-Facebook and the Census Household Pulse surveys. This estimation error may have more impact for smaller traditional surveys, such as Axios-Ipsos' survey, an issue we will investigate in subsequent work. 

We compute $\hat{\rho}_{Y,R_{\text{w}}}$ by using the benchmark $\bar Y_{_N}$, namely,  via solving Equation \eqref{eq:weiden} for $\hat{\rho}_{Y,R_{\text{w}}}$,
\begin{equation}\label{eq:werho}
   \hat\rho_{Y, R_{\text{w}}} =  \frac{Z_w}{\sqrt{N}},  \quad{\mbox{where}}\quad  Z_w= \frac{\bar{Y}_\text{w} - \bar{Y}_{_N}}  {\sqrt{\frac{1 - f_\text{w}}{n_\text{w}}} \sigma_Y}.
\end{equation}
We introduce this notation $Z_w$ because it is the quantity that determines the well-known survey efficiency measure, the so-called \textit{design effect}, which is the variance of  $Z_w$ for a probabilistic sampling design \cite{Kish1965} (when we assume the weights are fixed). For the more general setting where $\bar Y_w$ may be biased, we replace the variance by MSE, and hence the bias-adjusted design effect $D_e=E[Z_w^2]$, which is the MSE relative to the benchmark measured in the unit of the variance of an average from a simple random sample of size $n_w$.  Hence $D_I\equiv E[\hat\rho_{Y,R_{\text{w}}}^2]$, which was termed as the \textit{data defect index}\cite{Meng2018}, is simply the bias-adjusted design effect \textit{per unit}, because $D_I=D_e/N$.  

Furthermore, because $Z_w$ is the standardized actual error, it captures any kind of error inherited in $\bar Y_w$. This observation is important because when $Y$ is subject to measurement errors, $Z_w/\sqrt{N}$ no longer has the simple interpretation as a correlation.  But because we estimate $D_I$ by $Z_w^2/N$ directly, our effective sample size calculation is still valid even when Equation \eqref{eq:weiden} does not hold.

\section*{Asymptotic behavior of \textit{ddc}}
\label{sec:ddc-logit}

As shown in Meng\cite{Meng2018}, for any probabilistic sample without selection biases,  the \textit{ddc} is on the order of $1/\sqrt{N}$. Hence the magnitude of $\hrho$ (or $\hat\rho_{Y, R_{\text{w}}})$ is small enough to cancel out the impact of $\sqrt{N-n}$ (or $\sqrt{N-n_w}$) in the data scarcity term on the actual error, as seen in Equation \eqref{eq:meng_identity_math} (or Equation  \eqref{eq:weiden}). However, when a sample is unrepresentative, e.g.~when those with $Y = 1$ are more likely to enter the dataset than those with $Y = 0$, then $\hrho$ can far exceed $1/\sqrt{N}$ in magnitude. In this case, error will increase with $\sqrt{N}$ for a fixed \textit{ddc} and growing population size $N$ (Equation \eqref{eq:meng_identity_math}). This result may be counter-intuitive in the traditional survey statistics framework, which often considers how error changes as sample size $n$ grows. The \textit{ddc} framework considers a more general setup, taking into account individual response behavior, including  its impact on sample size itself. 

As an example of how response behavior can shape both total error and the number of respondents $n$, suppose individual response behavior is captured by a logistic regression model  
\begin{equation}
{\mbox{logit}} [\Pr(R=1|Y)] =  \alpha+\beta Y.
\end{equation}
This is a model for a response propensity score.
Its value is determined by $\alpha$, which drives the overall sampling fraction $f = n / N$, 
and by $\beta$, which controls how strongly $Y$ influences whether a participant will respond or not.

In this logit response model,  when $\beta\neq0$, $\hrho$ is determined by individual behavior, not by population size $N$. In Supplemental Information \ref{sec:logit-more}, we prove that \textit{ddc} cannot vanish as $N$ grows, nor can the observed sample size $n$ ever approach $0$ or $N$ for a given set of (finite and plausible) values of $\{\alpha, \beta\}$, because there will always be a non-trivial percentage of non-respondents. For example, an $f$ of 0.01 can be obtained under this model for either  ${\alpha = -0.46, \beta = 0}$ (no influence of individual behavior on response propensity), or for ${\alpha = -3.9, \beta = -4.84}$. However, despite the same $f$, the implied \emph{ddc} and consequently the MSE will differ. For example, the MSE for the former (no correlation with $Y$)  is 0.0004, while the MSE for the latter (a -4.84 coefficient on $Y$) is 0.242, over 600 times larger.

See Supplemental Information \ref{sec:heckman} for the connection between \textit{ddc} and a well-studied non-response model from econometrics, the Heckman selection model\cite{heckman1979sample}.

\section*{Population size in multi-stage sampling}
\label{sec:multipopulation}

We have shown that the asymptotic behavior of error depends on whether the data collection process is driven by individual response behavior or by survey design. The reality is often a mix of both. Consequently, the relevant ``population size'' $N$ depends on when and where the representativeness of the sample is destroyed, i.e., when the individual response behaviors come into play. Real-world surveys that are as complex as the three surveys we analyze here have multiple stages of sample selection.

Extended Data Table \ref{tab:census-ddc-stage} takes as an example the sampling stages of the Census Household Pulse, which has the most extensive set of documentation among the three surveys we analyze. As we have summarized (Table \ref{table:methodologies} and Extended Data Table \ref{table:methodologies-ED}), the Census Household Pulse (1) first defines the sampling frame as the reachable subset of the Master Address File, (2) takes a random sample of that population to prompt (send a survey questionnaire), and (3) waits for individuals to respond to that survey. Each of these stages reduces the desired data size, and the corresponding \textit{population size} is the intended sample size from the prior stage (in notation, $N_s = n_{s - 1}$, for $s = 2, 3$). For example, for stage 3, the population size $N_3$ is the size of the intended sample size $n_2$ from the second stage, i.e., the sampling stage, because only the sampled individuals have a chance to respond. 

Although all stages contribute to the grand \textit{ddc}, the stage that dominates is the \emph{first stage at which the representativeness of our sample is destroyed}---
whose size will be labeled as the \textit{dominating population size (dps)}---when the relevant population size decreases dramatically at each step. However, we must bear in mind that \textit{dps} refers to the worse case scenario, when biases accumulate, instead of (accidentally) cancel each other out. 

For example, if the 20 percent of the MAF excluded from the Census Household Pulse sampling frame (because they had no cell phone or email contact information) is not representative of the US Adult population, then the \textit{dps} is $N_1$, or 255 million adults contained in 144 million households.  Then the increase in bias for given \emph{ddc} is driven by the rate of $\sqrt{N_1}$ where $N_1 = 2.55 \times 10^8$ and is large indeed (with $\sqrt{2.5 \times 10^8} \approx 15,000$).  
In contrast, if the the sampling frame is representative of the target population and the outreach list is representative of the frame (and hence representative of the US adult population) but there is non-response bias, then \textit{dps} is $N_3=10^6$ and the impact \emph{ddc} is amplified by the square root of that number ($\sqrt{10^6} = 1,000$). In contrast, Axios-Ipsos reports a response rate of about $50\%$, and obtains a sample of $n = 1000$, so the \textit{dps} could be as small as $N_3 = 2000$ (with $\sqrt{2000}\approx 45$). 

This decomposition is why our comparison of the surveys is consistent with the \emph{Law of Large Populations} (estimation error increases with $\sqrt{N}$), \emph{even though all three surveys ultimately target the same US Adult Population}. Given our existing knowledge about online-offline populations\cite{ryan2016} and our analysis of Axios-Ipsos' small ``offline'' population, Census Household Pulse may suffer from unrepresentativeness at Stage 1 of Extended Data Table \ref{tab:census-ddc-stage} where $N =$ 255 million, and Delphi-Facebook may suffer from unrepresentativeness at the initial stage of starting from the Facebook User Base.  In contrast, the main source of unrepresentativeness for Axios-Ipsos maybe at a later stage where the relevant population size is orders of magnitude smaller.

\section*{CDC estimates of vaccination rates}

The CDC benchmark data used in our analysis was downloaded from the CDC's COVID data tracker \cite{cdc}. We employ the cumulative count of people who have received at least one dose of COVID-19 vaccine reported in the ``Vaccination Trends'' tab. This data set contains vaccine uptake counts for all US residents (not only adults). However, the surveys of interest estimate vaccine uptake among adults. The CDC receives age-group-specific data on vaccine uptake from all states except for Texas on a daily basis, which is also reported cumulatively over time.

Therefore, we must impute the number of adults who have received at least one dose on each day. We assume Texas is exchangeable with the rest of the states in terms of the age distribution for vaccine uptake. Under this assumption, for each day, we use the age group vaccine uptake data from all states except for Texas to calculate the proportion of cumulative vaccine recipients who are 18 or older, then we multiply that number by the total number of {people} who have had at least one dose to estimate the number of US \emph{adults} who have received at least one dose. 

The CDC performs a similar imputation for the 18+ numbers reported in their COVID data tracker. However the CDC's imputed 18+ number is available only as a snapshot and not a historical time series, hence the need for our imputation. See Supplemental Information \href{sec:SM-benchmark} for details of the imputation implementation.

\section*{Additional survey methodology}
\label{sec:SM-data-background}

The Census Household Pulse and Delphi-Facebook surveys are the first of their kind for each organization, while Ipsos has maintained their online panel for 12 years.

\paragraph{Question wording} 
All three surveys ask whether respondents have received a COVID-19 vaccine. See Extended Data Table \ref{table:methodologies-ED}. Delphi-Facebook and Census Household Pulse ask similar questions (``Have you had / received a  COVID-19 vaccination / vaccine?''). Axios-Ipsos asks ``Do you personally know anyone who has already received the COVID-19 vaccine?,'' and respondents are given response options including ``Yes, I have received the vaccine.'' The Axios-Ipsos question wording might pressure respondents to conform to their communities' modal behavior and thus misreport their true vaccination status, or may induce acquiescence bias from the multiple ``yes'' options presented. This pressure may exist both in high- and low-vaccination communities, so its net impact on Axios-Ipsos' results is unclear. Nonetheless, Axios-Ipsos' question wording does differ from that of the other two surveys, and may contribute the observed differences in estimates of vaccine uptake across surveys.

\paragraph{Population of Interest}

All three surveys target US adult population, but with different sampling and weighting schemes.
Household Pulse sets the denominator of their percentages as the household civilian, non-institutionalized population in the United States of 18 years of age or older, excluding Puerto Rico or the island areas. Axios-Ipsos designs samples to representative of the US general adult population 18 or older. For Facebook, the US target population reported in weekly contingency tables is the US adult population, excluding Puerto Rico and other US territories. For the CDC Benchmark, we define the denominator as the US 18+ population, excluding Puerto Rico and other US territories. To estimate the size of the total US population, we use the US Census Bureau Annual Estimates of the Resident Population for the United States and Puerto Rico, 2019 \cite{censuspopest}. This is also what the CDC uses as the denominator in calculating rates and percentages of the US population \cite{cdcvaxdemodata}.

Axios-Ipsos and Delphi-Facebook generate target distributions of the US adult population using the Current Population Survey (CPS), March Supplement, from 2019 and 2018, respectively. Census Household Pulse uses a combination of 2018 1-year American Community Survey (ACS) estimates and the Census Bureau's Population Estimates Program (PEP) from July 2020. Both the CPS and ACS are well-established large surveys by the Census and the choice between them is largely inconsequential.

\paragraph{Axios-Ipsos Data}
The Axios-Ipsos Coronavirus tracker is an ongoing, bi-weekly tracker intended to measure attitudes towards COVID-19 of adults in the US. The tracker has been running since March 13, 2020 and has released results from 45 waves as of May 28, 2021. Each wave generally runs over a period of 4 days. The Axios-Ipsos data used in this analysis was scraped from the topline PDF reports released on the Ipsos website \cite{axiosipsos}. The PDF reports also contain Ipsos' design effects, which we have confirmed are calculated as 1 plus the variance of the (scaled) weights. 

\paragraph{Census Household Pulse Data}
The Census Household Pulse is an experimental product of the US Census Bureau in collaboration with eleven other federal statistical agencies. We use the point estimates presented in Data Tables, as well as the standard errors calculated by the Census Bureau using replicate weights. The design effects are not reported, however we can calculate it as $1 + CV^2_{\text{w}}$, where $CV_{\text{w}}$ is the coefficient of variation  of the individual-level weights included in the microdata\cite{Kish1965}.

\paragraph{Delphi-Facebook COVID symptom survey}
The Delphi-Facebook COVID symptom survey is an ongoing survey collaboration between Facebook, the Delphi Group at Carnegie Mellon University (CMU), and the University of Maryland \cite{Barkay2020}. The survey is intended to track COVID-like symptoms over time in the US and in over 200 countries. We use only the US data in this analysis. The study recruits respondents using a daily stratified random samples recruiting a cross-section of Facebook Active Users. New respondents are obtained each day, and aggregates are reported publicly on weekly and monthly frequencies. The Delphi-Facebook data used here was downloaded directly from CMU's repository for weekly contingency tables with point estimates and standard errors. 

\newpage
\bibliography{bib}
\bibliographystyle{unsrtnat} 

\newpage

\section*{Ethical compliance}

According to HRA decision tools (\url{http://www.hra-decisiontools.org.uk/research/}), our study is considered Research, and according to the NHS REC review tool (\url{http://www.hra-decisiontools.org.uk/ethics/}), we do not need NHS Research Ethics Committee (REC) review, as we used only (1) publicly available, (2) anonymized, and (3) aggregated data outside of clinical settings.

\section*{Data availability}
Raw data is deposited in the Harvard Dataverse, at \url{https://doi.org/10.7910/DVN/GKBUUK}. Data was collected from publicly available repositories of survey data by downloading it directly or using APIs.

\section*{Code availability}
Code to replicate the findings is available in the repository \url{https://github.com/vcbradley/ddc-vaccine-US}. The main decomposition of the \emph{ddc} is available on the package ``ddi'' from the Comprehensive R Archive Network (CRAN).

\section*{Acknowledgments}
We thank Frauke Kreuter, Alex Reinhart, and the Delphi Group at Carnegie Mellon University, Facebook's Demography and Survey Science group; Frances Barlas, Chris Jackson, Catherine Morris, Mallory Newall, and the Public Affairs team at Ipsos; and Jason Fields and Jennifer Hunter Childs at the US Census Bureau for productive conversations about their surveys. We further thank the Delphi Group at CMU for their help in computing weekly design effects for their survey, the Ipsos team for providing data on their ``offline'' respondents, and the CDC for responding to our questions. Susan Paddock, other participants at the JPSM 2021 lecture (delivered by Meng), and Steve Finch provided helpful comments, which we greatly appreciate. We thank the anonymous reviewers for their constructive comments, which substantially improved our work. We thank Ariel Edwards-Levy for a tweet which originally inspired our interest in this topic, and Rick Born for suggesting more intuitive terms used in Equation~\eqref{eq:meng_identity}. V.B. is funded by the University of Oxford's Clarendon Fund and the EPSRC and MRC through the OxWaSP CDT programme (EP/L016710/1). X-L. M acknowledges partial financial support by NSF.   S.F. acknowledges the support of the EPSRC (EP/V002910/1).

\paragraph*{Author contributions}
V.B.~and S.F.~conceived and formulated the research questions. V.B.~and S.K.~contributed equally to data analysis, writing, and visualization. X-L.M.~conceived and formulated the methodology. All authors contributed to methodology, writing, visualization, editing, and data analysis. S.F.~supervised the work.

\paragraph*{Competing Interests}
Authors have no competing interests, financial or otherwise.

\newpage
\begin{center}
{\LARGE \textbf{Extended Data}}
\end{center}

\setcounter{figure}{0}
\setcounter{table}{0}
\renewcommand{\figurename}{Extended Data Fig.}
\renewcommand{\tablename}{Extended Data Table}
\addtocontents{toc}{\protect\setcounter{tocdepth}{2}}

\begin{figure}[p]
    \centering
    \includegraphics[width =0.85\textwidth]{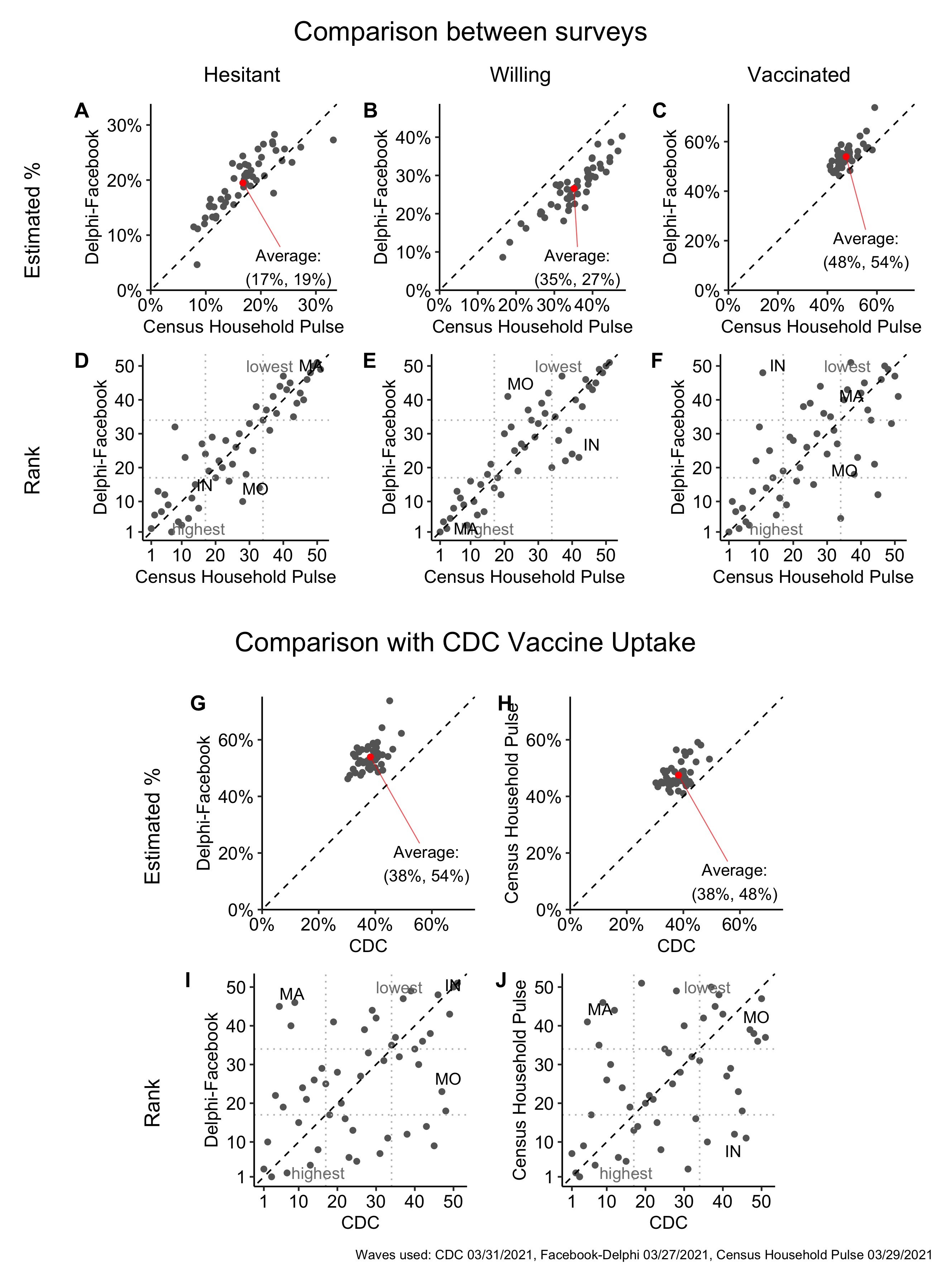}
    \caption{\textbf{Comparisons of state-level vaccine uptake, hesitancy, and willingness across surveys and the CDC: March 2021} Comparison of state-level point estimates (A-C) and rankings (D-F) for vaccine hesitancy, willingness, and uptake from Delphi-Facebook, and Census Household Pulse. Dotted black lines show agreement and red points show the average of 50 states. Panels G-J compare state-level point estimates and rankings for the same survey waves to CDC benchmark estimates from March 31, 2021. The Delphi-Facebook data is from the week ending  March 27, 2021 and the Census Household Pulse is the wave ending March 29, 2021.
    \label{fig:hesitancy-state-march}
    }
     \thisfloatpagestyle{empty}
\end{figure}

\begin{figure}[p]
    \centering
    \includegraphics[width =0.85\textwidth]{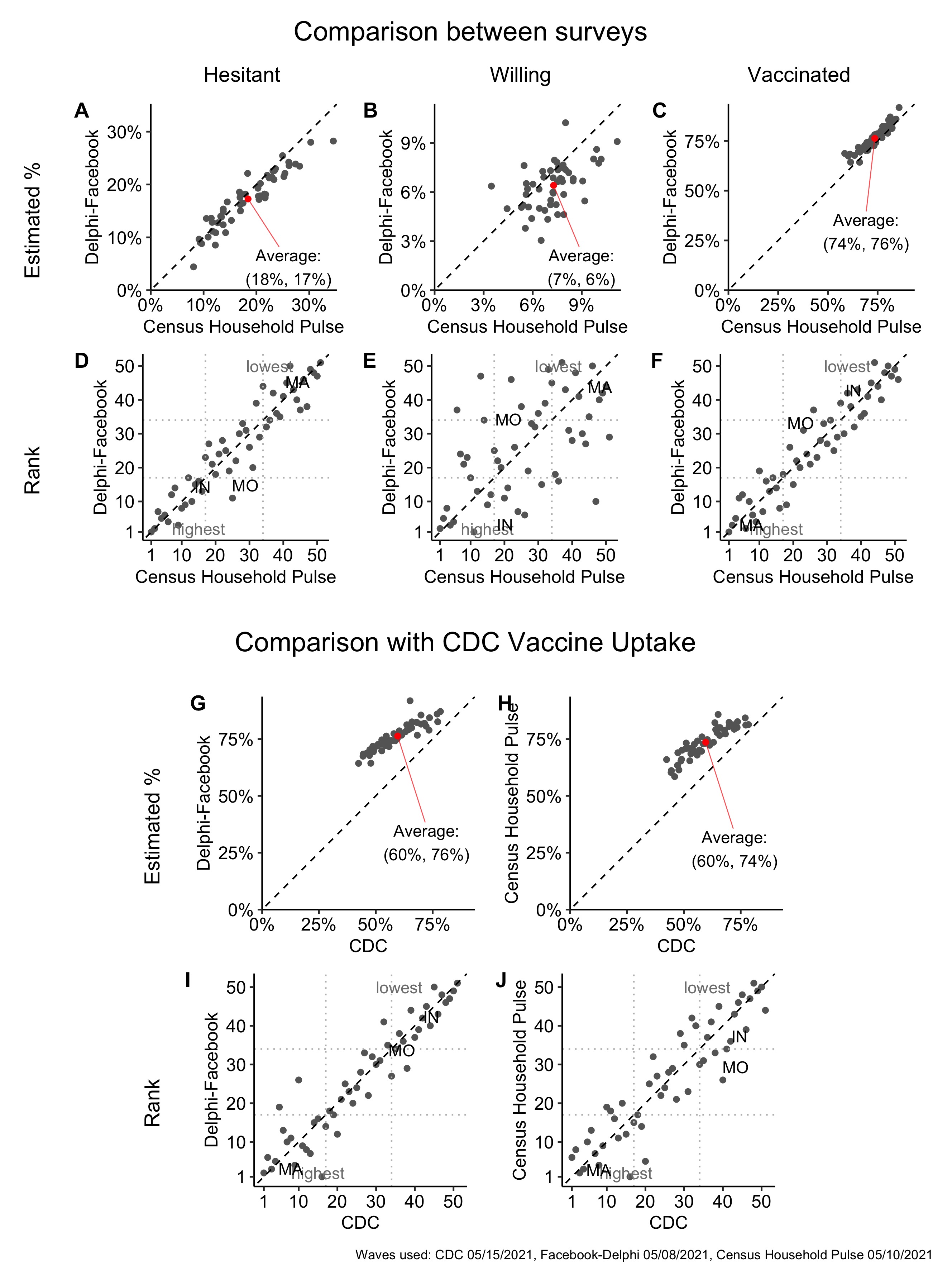}
    \caption{\textbf{Comparisons of state-level vaccine uptake, hesitancy, and willingness across surveys and the CDC: May 2021.} Comparison of state-level point estimates (A-C) and rankings (D-F) for vaccine hesitancy, willingness, and uptake from Delphi-Facebook, and Census' Household Pulse. Dotted black lines show agreement and red points show the average of 50 states. Panels G-J compare state-level point estimates and rankings for the same survey waves to CDC benchmark estimates from May 15, 2021. The Delphi-Facebook data is from the week ending  May 8, 2021 and the Census Household Pulse is the wave ending May 10, 2021.
    \label{fig:hesitancy-state-may}
    }
     \thisfloatpagestyle{empty}
\end{figure}

\begin{table}[p]
\footnotesize
    \centering
    \medskip
    \begin{tabular}{L{2.4cm}  L{4.15cm}  L{4.15cm}  L{4.15cm} }
    \toprule
        & \textcolor{ax}{\textbf{Axios-Ipsos}}   & \textcolor{hp}{\textbf{Census Household Pulse}} & \textcolor{fb}{\textbf{Delphi-Facebook}}  \\ \midrule
        \textbf{Purpose} & Measure national attitudes toward COVID-19 & Sub-national social and economic impact of COVID-19 & Fine-grained COVID-19 symptom surveillance \\ \hline
        \textbf{Target Pop.} & 18+ US general pop & 18+ US general pop &  18+ US general pop\\ \hline
        \textbf{Length of wave} & 4 days, conducted weekly & 2 weeks & Daily cross-section samples, reported weekly \\ \hline 
        \textbf{Average participation rate among invitees} & \multicolumn{1}{l}{50\%} & \multicolumn{1}{l}{6-8\%} & \multicolumn{1}{l}{1\%} \\ \midrule
        \textbf{Sampling design} & Inverse response propensity sampling & Systematic sample of households, adjusted for a projected response rates & Unequal-probability stratified random samples \\ \hline
        \textbf{Hesitancy / Willingness question} & ``How likely, if at all, are you to get the first generation COVID-19 vaccine, as soon as it's available'' & ``Once a vaccine preventing COVID-19 is available to you, would you...'' & ``If a vaccine to prevent COVID-19 were offered to you today, would you choose to get vaccinated?'' \\ \hline
        \textbf{Vaccine hesitancy responses} & ``Not very / at all likely'' & ``Definitely/Probably NOT get a vaccine'' or ``Unsure'' & ``No, definitely/probably not'' \\ \hline
        \textbf{Languages} & English and Spanish & English and Spanish & English, Spanish, Brazilian Portuguese, Vietnamese, French, and Chinese \\ \hline
        \textbf{Report MoE or design effect} & Both & Report standard errors for estimates from replicate weights & Report standard errors for estimates (does not include variance from weighting) \\ \hline
        \textbf{Sources for demographic benchmarks} & 2019 CPS March Supplement, party ID from recent ABC/WaPo polls & 2018 ACS, 1-year estimates & 2018 CPS March Supplement \\
        \bottomrule
    \end{tabular}
     \thisfloatpagestyle{empty}
         \caption{\textbf{Methodologies of Axios-Ipsos, Census Household Pulse, and Delphi-Facebook studies.} Supplements information in Table \ref{table:methodologies}.}
    \label{table:methodologies-ED}
\end{table}

\newpage

\begin{figure}[tb]
    \centering
    \includegraphics[width =0.8\linewidth]{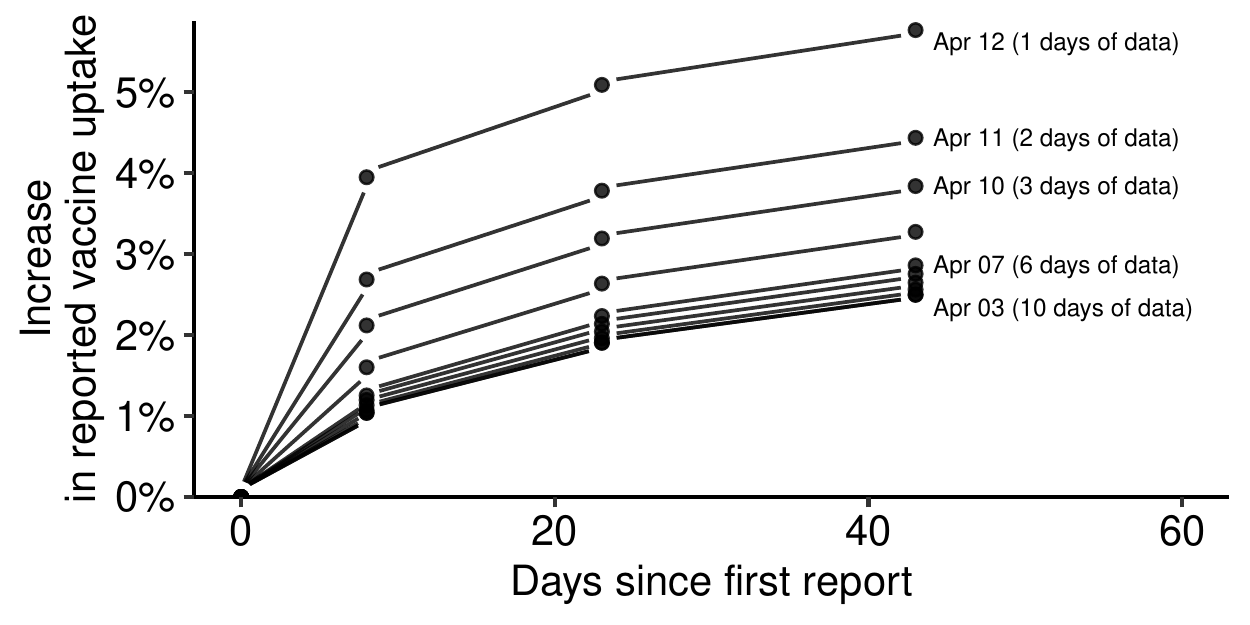}
    \caption{Retroactive adjustment of CDC vaccine uptake figures for April 3-12, 2021, over the 90 days from April 12. Increase is shown as a percentage of the vaccine uptake reported on April 12. Most of the retroactive increases in reported estimates appear to occur in the first 10 days after an estimate is first reported. By about 40 days after the initial estimates for a particular day are reported, the upward adjustment plateaus at around 5-6\% of the initial estimate. We use this analysis to guide the choice of 5\% and 10\% error in the CDC benchmark for our robustness checks.}
    \label{fig:benchmark-change}
\end{figure}

\newpage
\begin{figure}[thbp]
    \centering
        \includegraphics[width = \linewidth]{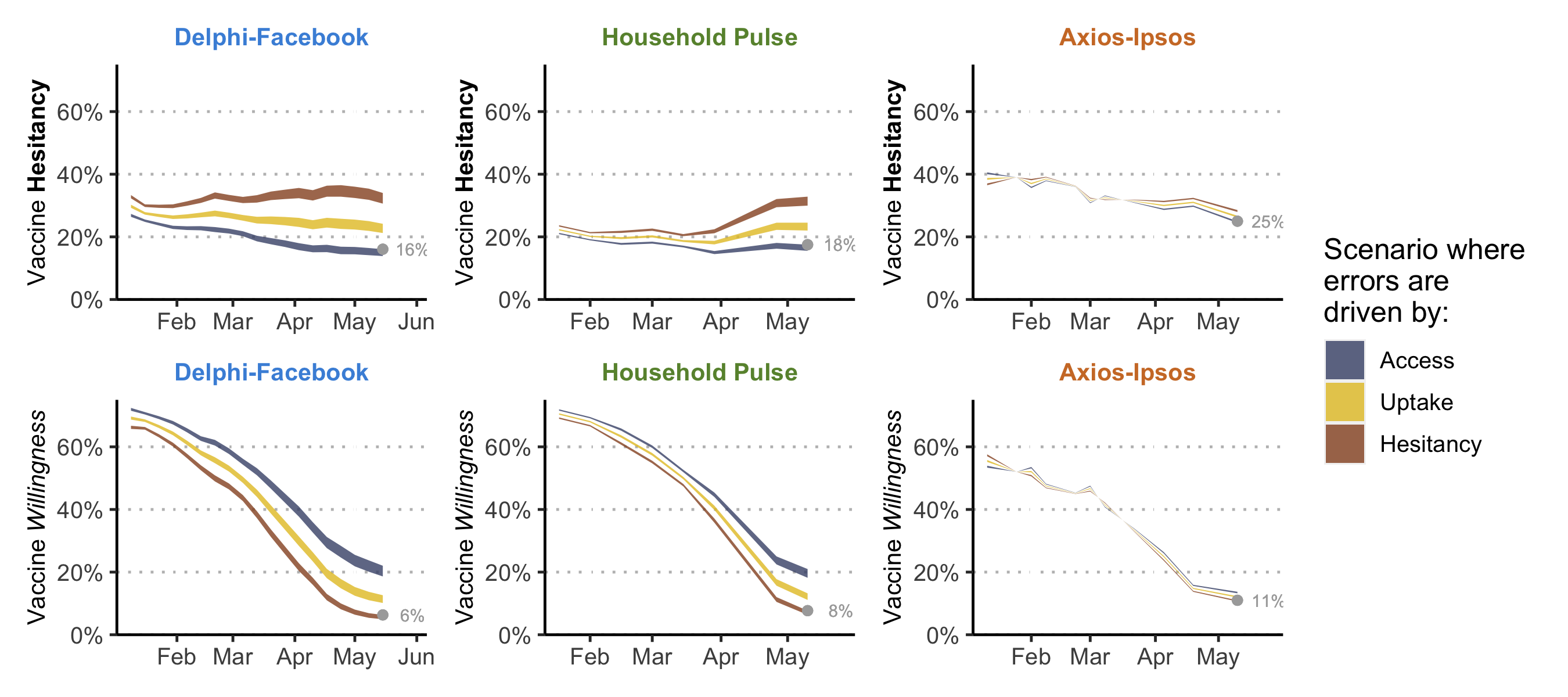}
     \thisfloatpagestyle{empty}
      \caption{\textbf{Revised estimates of hesitancy and willingness after accounting for survey errors for vaccination uptake.} The gray point shows the reported value at the last point of the time series. Each line shows a different scenario for what might be driving the error in uptake estimate, derived using hypothetical {\em ddc} values for willingness and hesitancy based on the observed {\em ddc} value for uptake. \textit{Access} scenario: willingness suffers from at least as much, if not more, bias than uptake. \textit{Hesitancy} scenario: hesitancy suffers from at least as much, if not more, bias than uptake. \textit{Uptake} scenario: the error is split roughly equally between hesitancy and willingness.
      See Supplementary Information \ref{sec:SM-scenarios} for more details.
      }
          \label{fig:scenario-allsurveys}
\end{figure}

\begin{figure}[h]
\includegraphics[width = \linewidth]{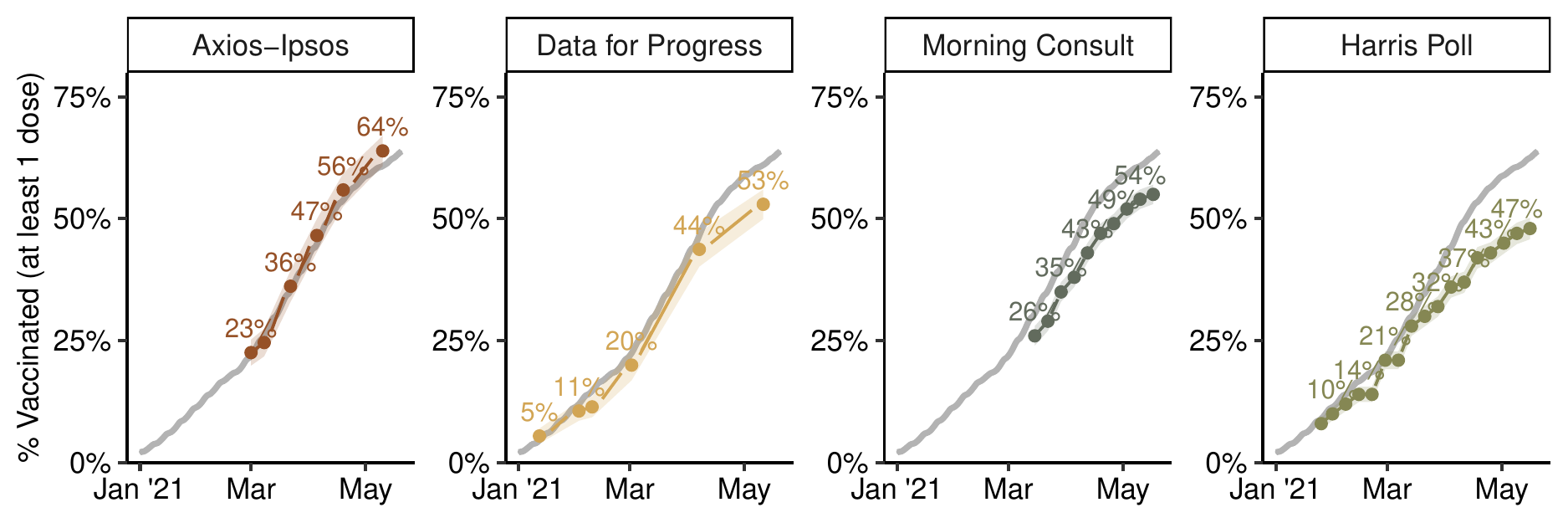}
\caption{\textbf{Vaccination Rates compared with CDC benchmark for four online polls}. Ribbons indicate traditional 95 percent confidence intervals which are twice the standard error reported by the poll. Data for Progress asks
``As of today, have you been vaccinated for Covid-19?''; Morning Consult asks ``Have you gotten the vaccine, or not?''; Harris Poll asks ``Which of the following best describes your mindset when it comes to getting the COVID-19 vaccine when it becomes available to you?''. See the Supplementary Information \ref{sec:SM-morepolls} for more details on each survey and discussion of differences.  Gray line is the CDC benchmark.}
\label{fig:morepolls}
\end{figure}

\begin{figure}[!hb]
    \centering
    \includegraphics[width = 0.7\linewidth]{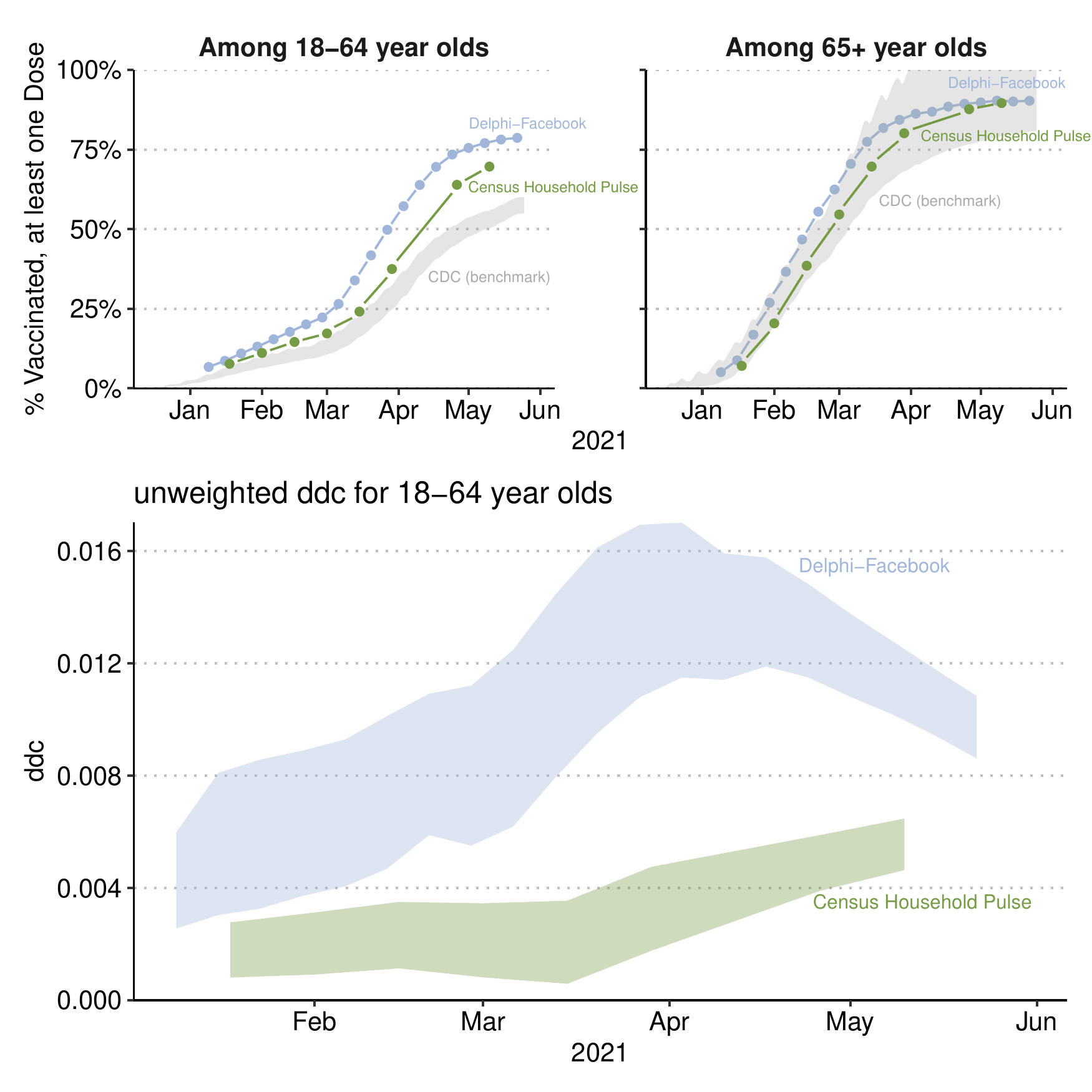}
    \caption{\textbf{Survey error by Age Group (18-64 year-olds, and those 65 and over)}. \textbf{a.} Estimates of vaccine uptake from Delphi-Facebook (blue) and Census Household Pulse (green) for each 18-64 year-olds (left) and those 65 or older (right). Bounds on the CDC's estimate of vaccine uptake for those groups are shown in gray. The CDC receives vaccination-by-age data only from some jurisdictions. We do know, however, the total number of vaccinations in the U.S. Therefore, we calculate the bounds allocating all the vaccine doses for which age is unknown to either 18-64 or 65+. \textbf{b.} Unweighted \textit{ddc} for each Delphi-Facebook and Census Household Pulse calculated for the 18-64 group using the bounds on the CDC's estimates of uptake. \textit{ddc} for 65+ is not shown due to large uncertainty in the bounded CDC estimates of uptake. \label{fig:ddc_by-age}
    }
\end{figure}

\newpage

\FloatBarrier

\begin{table}[p]
    \centering
    \begin{tabular}[t]{rcccc}
    \toprule
    \multicolumn{1}{c}{ } & \multicolumn{2}{c}{Vaccinated} & \multicolumn{1}{c}{Hesitant}\\
    \cmidrule(l{3pt}r{3pt}){2-3} \cmidrule(l{3pt}r{3pt}){4-4} \cmidrule(l{3pt}r{3pt}){5-5}
     & Raw & Weighted & Weighted & Sample size\\
    \midrule
    \multicolumn{5}{l}{\textbf{Axios-Ipsos Survey}}\\
    only Offline Panelists & 19\% & 13\% & 64\% & 21
\\
    only Online Panelists & 43 & 37 & 30 & 974
\\\addlinespace
    with Ipsos Weights & 42 & 36 & 30 & 995
\\
    with Delphi-implied Weights & 42 & 37 & 29 & 995
\\\\
    \multicolumn{5}{l}{\textbf{Delphi-Facebook Survey}}\\
    with Delphi Weights & 42\% & 46\% & 37\% & 249,954
\\\addlinespace
\bottomrule
\end{tabular}
\caption{\textbf{Contribution of offline recruitment and weighting schemes to discrepancies between surveys}. A portion of each Axios-Ipsos wave is recruited from a population with no stable internet connection; Ipsos KnowledgePanel provides tablets to these respondents. In the Axios-Ipsos March 22 2020 wave, the offline panelists ($n = 21$) were 24 percentage points less likely to be vaccinated than online panelists ($n = 974$). Weighting the same Axios-Ipsos data ($n = 995$) to the age and gender target distribution implied by Delphi-Facebook's weights make the vaccination estimates higher by 1 percentage point. However, this number is still lower than Delphi-Facebook's (responses from March 14--20 2020, $n = 249,954$) own estimate of 46\%. During this time period, the CDC benchmark vaccination rate was 35.2\%. This suggests that the recruitment of offline respondents and different weighting schemes each explains only a small portion of the discrepancy between the two data sources.}
\label{tab:ax-offline}
\end{table}

\newpage

\FloatBarrier

\begin{table}[p]
    \centering
    \small
    \begin{tabular}{lcccc}
    \toprule
        \textbf{Stage} $\bm{s}$ &  \textbf{Population} $\bm{N_{s}}$ &  \textbf{Sampling Process} $\longrightarrow$ & \textbf{Data} $\bm{n_s}$ & $\bm{n_s}/\bm{N_s}$  \\ \midrule
        1. {Define frame} & 144 m hh & Subset to reachable address &  116 m hh & 80\% \\
        2. {Decide outreach list} & 116 m hh & Random sample & 1 m adults & $1\%$ \\
        3. {Individual behavior} & 1 m adults & Individual responds (or doesn't) & 75,000 adults & 7\%\\ \midrule
        Final & $\sim$ 255 m adults & & 75,000 adults & 0.03\% \\\bottomrule
    \end{tabular}
    \caption{\textbf{Example of multi-stage population selection}. The \emph{Law of Large Populations} described in Methods section ``\nameref{sec:multipopulation}'' shows that the population size at the sampling stage where simple random sampling breaks down will dominate the error. This table explains these stages with a concrete example, using the Census Household Pulse.  Population and sample sizes for three stages (stage number denoted $s \in \{1,2,3\}$) of sampling  of the Census Household Pulse survey data collection process. Approximate sample sizes based on the March 24, 2021 wave. ``m'' stands for millions and ``hh'' stands for household.   
    The final row compares the total adult population in the US (255 million adults, made up of 144 million households) to the sample size in one wave of the household pulse. For the purpose of illustration, we have ignored the impact of unequal sampling probabilities on the sample sizes at each stage.
    }
    \label{tab:census-ddc-stage}
\end{table}

\FloatBarrier

\appendix
\setcounter{figure}{0}
\setcounter{table}{0}
\renewcommand{\figurename}{Fig.}
\renewcommand{\tablename}{Table}
\renewcommand\thefigure{S\arabic{figure}}
\renewcommand\thetable{S\arabic{table}}
\addtocontents{toc}{\protect\setcounter{tocdepth}{2}}

\newpage

\setcounter{page}{1}

\begin{center}
{\LARGE \textbf{Supplementary Information}}
\end{center}

\vspace{0.1in}

\setcounter{secnumdepth}{2}
\setcounter{tocdepth}{2}
\begin{center}
\begin{minipage}{0.8\linewidth}
\tableofcontents
\end{minipage}
\end{center}

\newpage
\FloatBarrier

\newpage

\section{Additional Details about Data Sources}

\subsection{Total Population}

The CDC vaccination data includes vaccines administered in Puerto Rico. As of June 9, 2021, approximately 1.6 million adults have received at least one dose, just under 1\% of the national total (164,576,933). We use the CDC's reported national total that includes Puerto Rico (we do not have a reliable state-level time series of vaccine uptake), but we use a denominator that \emph{does not} include Puerto Rico. This means that the CDC's estimate of vaccine uptake used here may be slightly \emph{overestimating} the true proportion of the US (non-Puerto Rico) adult population that has received at least one dose by about 1\%, which would make the observed \emph{ddc} for Delphi-Facebook and Census Household Pulse and \emph{underestimate} of the truth. However, this 1\% error is well within the benchmark uncertainty scenarios presented with our results.

\subsection{CDC Imputation and Uncertainty}
\label{sec:SM-benchmark}

The CDC does release state-level snapshots of vaccine uptake each day. These have been scraped and released publicly by Our World In Data \cite{owid}. These state-level numbers are not historically-updated as new reports of vaccines administered on previous days are reported to the CDC, so they underestimate the true rate of state-level vaccine uptake on any given day. These data are used only to motivate the inaccuracies of the state-level rank orders implied by vaccine uptake estimates from Delphi-Facebook and Census Household Pulse; hence they are not used to calculate $ddc$.

To inform our CDC benchmark uncertainty scenarios, we examined changes in vaccine uptake rates reported by the CDC over time.  We downloaded versions of the CDC's cumulative vaccine uptake estimates that are updated retroactively as new reports of vaccinations are received on April 12, April 21, May 5, and May 26. This allowed us to examine how much the CDC's estimates of vaccine uptake for a particular day change as new reports are received. Extended Data Fig.~\ref{fig:benchmark-change} compares the estimates of cumulative vaccine uptake for April 3-12, 2021 reported on April 12, 2021 to estimates for those same dates reported on subsequent dates. The top line shows that the cumulative vaccine uptake estimate for April 12, 2021 is, over the next month and a half, adjusted upwards by approximately 6\% of the original estimate reported on April 12, 2021 . The estimate of vaccine uptake for April 11, reported on April 12, is only further adjusted upward by approximately 4\% over the next 45 days. There is little apparent difference in the amount by which estimates from April 3-8 are adjusted upwards after 45 days, indicating that most of the adjustment occurred in the first 4 days after the initial report, which is consistent with the CDC's findings \cite{cdc}. There is still some adjustment that occurs past day 5; after 45 additional days, estimates are adjusted upwards by an additional 2\%.

There are many caveats to this analysis of CDC benchmark under-reporting, including that it depends on snapshots of data collected at inconsistent intervals, and that we mainly examine a particular window of time, April 3-12, so our results may not generalize to other windows of time. This is plausible for a number of reasons including changes to CDC reporting systems and procedures after the start of the mass vaccination program, or due to the fact that true underlying vaccine uptake is monotonically increasing over time. It is also plausible, if not likely, that the reporting delays are correlated with vaccine providers which are in turn correlated with the population receiving vaccines at a given time. As the underlying population receiving vaccines changes, so would the severity of reporting delays.

We use these results to inform our choice of benchmark uncertainty scenarios: 5\% and 10\%. The benchmark error is incorporated into our analysis by adjusting the benchmark estimates each day up or down by 5\% or 10\% (i.e. multiplying the CDC's reported estimate by 0.9, 0.95, 1.05, and 1.1). We then calculate \textit{ddc} on each day for each error scenario, as well as for the CDC reported point estimate.

However, the benchmark data that we use here \textit{has} been retroactively-adjusted as new reports of vaccine administration are received, so that the scenarios we consider are in addition to the initial reporting lag which has already been accounted for. These scenarios are intended only to demonstrate the robustness of our findings to plausible latent error in the benchmark data rather than to suggest that those scenarios are at all likely. To fully account for errors in the CDC benchmark would require a close collaboration with the CDC, and to have access to its historical information and methodologies on addressing issues such as never-reporting, as occurred when reporting AIDS status \cite{tu1993aids,bouman2005bayesian}.

\subsection{Availability of Survey Microdata}

Both Axios-Ipsos and Census Household Pulse release microdata publicly. Facebook also releases microdata to institutions that have signed Data Use Agreements. In view of the timeliness of our study, and to keep all three surveys on as equal a footing as possible,  we used the aggregated results released by all three surveys rather than their microdata. 

In all surveys, data collection happens over a multi-day period (or multi-week in the case of the Census Household Pulse). We calculate error for each survey wave with respect to the CDC-reported proportion of the population vaccinated up to and including the end date of each wave. Some respondents will have actually responded days (or weeks) before the date on which the estimate was released, when the true rate of vaccine uptake was lower. We use the end date instead of a mid-point as we do not have good data on how respondents are distributed over the response window. However, this means that the error we report may \textit{underestimate} the true error in each survey, particularly those with longer fielding and reporting windows.

\subsection{Census Household Pulse}

The Census Household Pulse is administered by the the Bureau of Labor Statistics (BLS); the Bureau of Transportation Statistics (BTS); the Centers for Disease Control and Prevention (CDC); Department of Defense (DOD); the Department of Housing and Urban Development (HUD);  Maternal and Child Health Bureau (MCHB); the National Center for Education Statistics (NCES); the National Center for Health Statistics (NCHS); the National Institute for Occupational Safety and Health (NIOSH); the Social Security Administration (SSA); and the USDA Economic Research Service (ERS) (\url{https://www.census.gov/programs-surveys/household-pulse-survey.html}, visited June 5, 2021). Each wave since August 2020 fields over a 13-day time window. All data used in this analysis is publicly available on the US Census website. 

The Census Household Pulse changed the question used to gauge vaccine willingness and hesitancy beginning with wave 27 (the most recent wave used in this analysis), to add a response option for respondents who are ``unsure'' if they will receive a COVID vaccine when they become eligible. Approximately 6.6\% of all respondents reported being ``unsure'' in wave 27, and were coded as ``vaccine hesitant'' rather than ``willing.''

\subsection{Delphi-Facebook}

Facebook performs inverse propensity weighting on responses, but the reported standard errors do not include variance increases from weighting, and no estimates of design effects are released publicly. We are therefore grateful to the CMU team for providing us with estimated weekly design effects for all weeks through April 2021. The design effects are quite consistent across 2021 waves (Mean: $1.48$, 95\% CI: $1.48-1.49$), so we mean-impute the design effects for May waves.

\newpage
\section{Asymptotic Properties of \textit{ddc}}
\label{sec:SM-ddc-more}

Here we lay out the formal results underlying the interpretation of our empirical decomposition of total error into \emph{ddc}. The first section explains how individual response behavior drives $\hrho$ and sampling rate $f = n/ N$. The second section describes why the relevant population size $N$ differs between surveys of the same target population when the data collection process involves  multiple processes.  This clarifies the key distinction with the classic probabilistic sampling framework, and how our results are consistent with the \emph{Law of Large Populations}\cite{Meng2018}.

\subsection{The Role of Individual Response Behavior}
\label{sec:logit-more}

In the Methods ``\nameref{sec:ddc-logit}'', we considered a logit model of the propensity score to assert that the \textit{ddc} $\hrho$ will not vanish with the population size $N$, regardless of how large $N$ is. Here we provide the mathematical proof of this assertion.  First, recall that the probability calculation involving $Y$ is with respect to its finite population $\{Y_i, i=1,\ldots, N\}$, we have 
$\Pr(Y=1)=\bar Y_{_N}$. Therefore, when the individual response model
$$\Pr(R=1|Y) =  \frac{e^{\alpha+\beta Y}}{1+e^{\alpha+\beta Y}}$$ 
is applicable to the entire finite population (e.g., a social media platform is open to everyone, at least in theory), 
we have that, as $N\to\infty$, the fraction of observations
\begin{align}\label{eq:fmar}
    f \rightarrow (1-\mu)\frac{e^{\alpha}}{1+e^{\alpha}} +\mu\frac{e^{\alpha+\beta}}{1+e^{\alpha+\beta}}=: p,
\end{align} 
where $\mu \in (0, 1)$ denotes the limit of $\bar Y_{_N}$ as $N$ increase to infinity. Here we assume such a limit exists, and it is not a trivial one (that is, $\mu$ stays away from 0 or 1).  Consequently, $p\in(0,1)$, i.e. it also stays away from 0 or 1, since it is a convex combination of $\frac{e^{\alpha}}{1+e^{\alpha}}$ and $\frac{e^{\alpha+\beta}}{1+e^{\alpha+\beta}}$, both of which lie in $(0,1)$. 
This means that we cannot make the sample $n$ arbitrarily large (or small), such as approaching $N$, or even at a particular level, because it is controlled by the value of $\{\alpha, \beta\}$, which is determined by the individual response behavior (towards the specific question underlying $Y$).

Second,  because $\text{Cov}(Y, R) = \E(YR) - \E(Y)\E(R)= \Pr(R=1| Y=1)\Pr(Y=1)- \bar Y_{_N} f$, we have 
\begin{align}\label{eq:rhor}
    \hrho \rightarrow \left(\frac{e^{\alpha+\beta}}{1+e^{\alpha+\beta}}-\frac{e^{\alpha}}{1+e^{\alpha}}\right)\frac{\sqrt{\mu(1-\mu)}}{\sqrt{p(1-p)}}.
\end{align}
This implies that for any given value of $\{\alpha, \beta\}$,
$\hrho$ will converge to a non-zero value $\rho$ as long as $\beta\not=0$, that is, as long as the propensity for response depends on $Y$ itself.  Consequently, the total error, relative to the standard error from simple random sampling (as a benchmark), denoted by $Z$,
\begin{align}\label{eq:zeror}
Z=: \frac{\bar Y_{_N} - \bar Y_{_N}}{\sqrt{(1-f)\sigma^2/n}} = \hrho \sqrt{N} 
\end{align}
goes to infinity with $N$ at the rate of $\rho \sqrt{N}$, a phenomenon that does not happen when $\beta=0$.

\subsection{Connection with the Heckman selection model}
\label{sec:heckman}

The goal of the Heckman selection model\cite{heckman1979sample} is to perform estimation in the case of non-response induced by censoring a latent variable. Specifically, let each member of the population be identified via a tuple of characteristics $(Y_{1i}, Y_{2i})$ which satisfy:
\begin{equation}
    \begin{split}
        Y_{1i} = X_{1i} \beta_1 + U_{1i} \\
        Y_{2i} = X_{2i} \beta_2 + U_{2i},
    \end{split}
\end{equation}
where the tuples of $U_i$ are identically and independently distributed multivariate Normal noise:
\begin{equation}\label{eq:heckman_errorterm}
    \begin{pmatrix} U_{1i} \\ U_{2i} \end{pmatrix} \sim N \left( \mathbf{0}, \begin{pmatrix} \sigma_{1}^2 & r\sigma_1\sigma_2\\ r\sigma_1\sigma_2 & \sigma_{2}^2 \end{pmatrix} \right)
\end{equation}
and the $\beta_j$'s are regression coefficients. We seek to estimate $\beta_1$, but observe data $Y_{1i}$ if and only if $Y_{2i} \geq 0$ (the predictors $X_{ji}$ are observed for all members of the population, however). In our framework, the response indicator is $R_i = I(Y_{2i} \geq 0)$.
The \textit{ddc} $\rho$ under the Heckman model (which is a theoretical model and hence this is a theoretical calculation) then is given by, using properties of the multivariate Normal,
\begin{align}\label{eq:heckman_ddc}
    \begin{split}
        \rho &=\text{Corr}(Y_{1i}, I(Y_{2i} \geq 0)) \\ &=r \cdot \frac{\phi(Z_i)}{\sqrt{\Phi(-Z_i)[1 - \Phi(-Z_i)]}},
    \end{split}
\end{align}
where $Z_i= -X_{2i} \beta_2/\sigma_2$.
Hence in this case the \textit{ddc} is a multiplier of the correlation $r$ in \eqref{eq:heckman_errorterm}, where the multiplier  factor $\lambda_i'$ resembles the inverse Mills ratio $\phi(Z_i) / \Phi(Z_i)$, where $\phi$ and $\Phi$ are respectively the PDF and CDF of the standard Normal $N(0, 1)$.

Intuitively, it makes sense for $\rho$ to be closely tied with $r$, since $r$ drives the selection bias. For example, if $r=0$, then $Y_2$ is independent from $Y_1$, and hence the sign of $Y_2$ will carry no information about $Y_1$. Therefore, for the purpose of estimating $\beta_1$, the data information is not distorted by having the sample inclusion determined by the sign of $Y_2$, when $r=0$. Hence $r=0$ must imply $\rho=0$, and vice versa. However,  $r$ alone is insufficient to capture the impact of the biased selection mechanism, since minimally the mean of $Y_2$, which
impacts the $Z$ term, would influence which portion of the data is more likely to be observed. 
The \textit{ddc} $\rho$ provides a metric to capture the overall effect. 

In conclusion, the \textit{ddc} framework is closely related to the framework for inferring the population mean under the Heckman selection model (corresponding to set $X_1=1$). The benefit of the Heckman selection model is that we can also estimate the selection mechanism itself from the observed data thanks to the distributional assumptions about the data generating mechanism.  The downside of course is that the validity of our results will depend on the reliability of the assumptions.  In contrast, \textit{ddc} makes no distributional assumptions about the data generating process, and hence it is broadly applicable.  However, there is no free lunch -- we cannot estimate \textit{ddc} without external information.
Nonetheless, it is a useful metric in the presence of a ground truth or plausible set of scenarios for the outcome of interest, such as  in our paper.

\newpage

\section{Additional Data Analyses}

\subsection{Estimates of hesitancy by demographic groups} \label{sec:SM-subgroup-emprirics}

We show estimates of our main outcomes by Education, and then by Race, in Table \ref{tab:outcome-by-demo}. The estimates vary by mode, but the rank ordering of a particular outcome within a single survey is roughly similar across surveys. The same estimates from Household Pulse  were already presented in Table \ref{tab:edu-nonrepresentative}.

\begin{table}[htb]
\caption{\textbf{Levels of Vaccination, Willingness, and Hesitancy, estimated by demographic group.} For each outcome, we estimate the same quantity from the three surveys.  Axios-Ipsos (denoted AP).
Census Household Pulse (denoted HP): wave ending March 29, 2021, $n$ = 76,068.  Delphi-Facebook (denoted FB): wave ending March 27, 2021, $n$ = 181,949.
These are the same waves as those in Table \ref{tab:edu-nonrepresentative}. 
Axios does not record a separate category for Asian Americans (they are lumped into ``Other'', so the values are left blank.}
\label{tab:outcome-by-demo}
\bigskip
\centering
\begin{tabular}{r ccc ccc ccc}\toprule
 & \multicolumn{3}{c}{\% Vaccinated} & \multicolumn{3}{c}{\% Willing} & \multicolumn{3}{c}{\% Hesitant}\\
 \cmidrule(lr){2-4} \cmidrule(lr){5-7}  \cmidrule(lr){8-10}
Education & AX & HP & FB & AX & HP & FB & AX & HP & FB\\\midrule
High School & 28\% & 39\% & 40\% & 32\% & 40\% & 35\% & 40\% & 21\% & 25\%
\\
Some College & 36 & 44 & 52 & 30 & 38 & 27 & 34 & 18 & 21
\\
4-Year College  & 36 & 54 & 62 & 45 & 36 & 26 & 19 & 10 & 12
\\
Post-Graduate  & 56 & 67 & 73 & 33 & 26 & 19 & 10 & 7 & 9
\\
     \bottomrule
\end{tabular}

\bigskip

\begin{tabular}{r ccc ccc ccc}\toprule
 & \multicolumn{3}{c}{\% Vaccinated} & \multicolumn{3}{c}{\% Willing} & \multicolumn{3}{c}{\% Hesitant}\\
 \cmidrule(lr){2-4} \cmidrule(lr){5-7}  \cmidrule(lr){8-10}
Race & AX & HP & FB & AX & HP & FB & AX & HP & FB\\\midrule
White & 40\% & 50\% & 59\% & 29\% & 33\% & 24\% & 30\% & 17\% & 17\%
\\
Black & 27 & 42 & 55 & 44 & 39 & 28 & 29 & 19 & 17
\\
Hispanic  & 26 & 38 & 45 & 39 & 48 & 39 & 36 & 14 & 16
\\
Asian  &  & 51 & 58 &  & 43 & 37 &  & 5 & 5
\\
     \bottomrule
\end{tabular}
\end{table}

\subsection{\emph{ddc} by age / eligibility status across time}

The CDC also releases vaccination rates by age groups, albeit not always in bins that overlap with the survey.  For overlapping bins (seniors and non-seniors) we can calculate \textit{ddc} specific to each group (Extended Data Fig.~\ref{fig:ddc_by-age}).

The CDC only receives vaccination data for age groups from certain jurisdictions, so is likely unrepresentative of the entire US adult population. Therefore, we calculate wide bounds for what the true proportion of each age group could be based on allocating the administered doses for which we do not have age information either entirely to seniors or entirely to non-seniors. When this allocation implies a vaccination rate of more than 100\% for that group, the remaining doses are allocated to the other age group. For example, if we know that on a particular day, X doses were administered to non-seniors, Y doses were administered to seniors, and Z doses were administered for which we have no age information, then the bounds for non-seniors are calculated as (X, X+Z) divided by the size of the non-senior US population. Similarly, the bounds for seniors were calculated as (Y, Y+Z) divided by the size of the senior US population.

These bounds do not incorporate any additional benchmark error, so may suffer from reporting delays or other systematic biases, and should be interpreted with caution. We do not show \textit{ddc} for the 65+ age group due to the large width of the conservative bounds which led to unreliable estimates.

\FloatBarrier

\subsection{Other online polls}
\label{sec:SM-morepolls}

Clearly surveys can and do go wrong regardless of their sizes.  Therefore, they key message of our analysis is \emph{not} that ``the smaller the better'', but rather that (1) quality matters far more than quantity, and (2) large surveys fail more drastically than small surveys when there is non-negligible \emph{ddc}. To highlight these points, we considered three more major online polls that ask vaccination status.

Figure \ref{fig:morepolls} shows how the estimated vaccination rate of Axios-Ipsos, Data for Progress, Morning Consult, and Harris Poll tracks the CDC benchmark. The poll that is perhaps most similar to Axios-Ipsos and provides enough documentation of their methods and data, Data for Progress, generated similar patterns as Axios-Ipsos. Their estimates tended to underestimate the vaccination rate by May, but did not suffer from overconfidence in its incorrect estimate. Data for Progress is an online-only panel run in the online vendor Lucid. 

\paragraph{Data for Progress} collects samples by the online vendor Lucid. Each wave can last up to a week and has a sample size of about $n = 1,000$. They ask:
\begin{quote}
    ``As you may know, vaccines for Covid-19 have now been approved by the Food and Drug Administration and are being offered to some individuals based on specific criteria. As of today, have you been vaccinated for Covid-19?''
    (1) ``Yes, I have received at least one Covid-19 vaccination shot,''
    (2) ``No, I have not received a Covid-19 vaccination shot.''
\end{quote}

\noindent Data for Progress' poststratification weighting weights to national numbers of ``gender, age, region, education, race, the interaction of education and race, and presidential vote ([2020 presidential vote]).''

\paragraph{Harris Poll} employs an online panel with an unspecified vendor. Their weekly COVID polls are about $n = 2,000$ per wave, covering three days. They  ask:

\begin{quote}
    ``Which of the following best describes your mindset when it comes to getting the COVID-19 vaccine when it becomes available to you?'' (1) ``I plan to go the first day I am able to'', (2) ``Whenever I get around to it'', (3) ``I will wait awhile and see'', (4) ``I will not get a COVID-19 vaccine'', and (5) ``I have already received a COVID-19 vaccine.''
\end{quote}
and the analysis here only takes the last option as an indicator for vaccine uptake. 

The  Harris Poll weights by a propensity score by their ``propensity to be online,'' and additionally poststratify for  ``age, sex, race/ethnicity, education, region, household size, employment, and household income'' to population benchmarks.

\paragraph{Morning Consult} employs their own online panel. They report a margin of 1 percentage point and a rough sample size of $n = 30,000$ per week (which corresponds to a wave). They ask:
\begin{quote}
    ``Have you gotten the vaccine, or not?'' (1)  ``Yes'',  (2) ``No, but I will get it in the future,'' (3) ``No, and I am not sure if I will get it in the future,'' and (4) ``No, and I do not plan to get it.''
\end{quote}

\noindent Morning Consult weights their survey data to ``a range of demographic factors, including age, race/ethnicity, gender, educational attainment, and region. State-level results were weighted separately to be representative of age, gender, race/ethnicity, education, home ownership and population density.''

\textbf{YouGov} is also a prominent online poll. However,  YouGov, unlike the other polls discussed here, investigated how their estimates track the CDC vaccination rate\cite{yougovwording}. Therefore, we do not compare it with the other polls here.  They found that the ``have you been vaccinated'' wording was more accurate than starting the question with ``will you be vaccinated?'' and including an ``already'' option,  which tended to underestimate the vaccination rate. Their A/B test confirmed the change in question wording caused a discrepancy of about 14 percentage points even in the same poll.

YouGov's A/B test provides some indication why Harris underestimates the vaccination rate. Note that Harris, unlike Data for Progress and our three surveys in the main text, uses the wording, ``when [the vaccine] becomes available to you.'' This is precisely the type of question wording that would underestimate vaccination rates, per YouGov. The underestimation of Morning Consult may be separately due to its questions not specifying ``at least one dose,'' thereby inducing a fraction of one-dose only respondents to not select ``Yes.'' We therefore suspect the underestimation of the Harris Poll is due to the question wording rather than something systematic about online polls.

\newpage
\FloatBarrier

\section{\emph{ddc}-based Scenario Analysis for Willingness and Hesitancy}\label{sec:SM-scenarios}

The main quantity of interest in the surveys examined here is not uptake, but rather willingness and hesitancy to accept a vaccine when it becomes available. Our analysis of \emph{ddc} of vaccine uptake cannot offer conclusive corrected estimates of willingness and hesitancy; however we propose \emph{ddc}-based scenarios that suggest plausible values of willingness and hesitancy given specific hypotheses about the mechanisms driving selection bias.

\subsection{Setting up scenarios} 
We adopt the following notation for the key random variables we wish to measure:
\begin{itemize}
    \item $V$ - did you receive a vaccine (``vaccination'')?
    \item $W$ - if no, will you receive a vaccine when available (``willingness'')? 
    \item $H = 1 - V - W$ - vaccine ``hesitancy''
\end{itemize}
Just as we have studied the data quality issue for estimating the vaccine uptake, we can apply the same framework to both $W$ and $H$.  Unlike uptake, however, we do not have CDC benchmarks for willingness or hesitancy. We only know that $V + H + W = 1$, and therefore that 
\begin{align*}
    \text{Cov}(R, V) + \text{Cov}(R, H) + \text{Cov}(R, W) = 0
\end{align*}
Re-expressing the covariances as correlation, and recognizing that $\text{Corr}(R,\cdot) = \rho_{R, \cdot}$, we obtain 
\begin{align*}
    \rho_{R,V}\cdot \sigma_V + \rho_{R,H}\cdot \sigma_H + \rho_{R,W}\cdot \sigma_W = 0
\end{align*}

It is well-known that for a Bernoulli random variable, its variance is rather stable around 0.25 unless its mean is close to 0 or 1. 
For simplicity, we then adopt the approximation that $\sigma^2_V \approx \sigma_H^2 \approx \sigma_W^2$. Consequently, we have
$$\rho_{R,V}+ \rho_{R,H} + \rho_{R,W} \approx 0$$

As we have estimated $ddc$ of vaccine uptake for each survey wave, we can further say that $\rho_{R,H} + \rho_{R, W} \approx  - \hat\rho_{R,V}$. However, we have no information to suggest how $\rho_{R,V}$ is decomposed into $ddc$ of hesitancy and willingness.  Therefore, we introduce a tuning parameter, $\lambda$, that allows us to control the relative weight given to each $\rho_{R,H}$ and $\rho_{R, W}$, such that
\begin{align*}
    -\rho_{R, H} = (1 - \lambda)  \hat\rho_{R,V}, \quad 
    -\rho_{R, W} = \lambda  \hat\rho_{R,V}
\end{align*}
The tuning parameter $\lambda$ may take on values greater than 1 and less than -1, which would indicate that the $ddc$ of either willingness or hesitancy is \textit{greater} in magnitude than that of uptake, or that selection bias is more extreme than that or vaccine uptake.

\subsection{Obtaining scenario estimates}

Once we postulate a particular value of \textit{ddc}, we can use Equation \ref{eq:weiden} to solve for the population quantity of interest, say $\bar H_N$.  
Specifically, given a postulated value of $\rho_{_{H,R_{\text{w}}}}=r$, we can calculate $\bar H_N$ as follows:
\begin{align}
    \overline{H}_\text{w} - \overline{H}_N & = \underbrace{r \cdot \sqrt{\frac{N-n_\text{w}}{n_\text{w}}}}_c \cdot \sqrt{\overline{H}_N (1 - \overline{H}_N)}. \label{eq:pre_square}
\end{align}
Squaring both sides and rearranging, we obtain:
\begin{align}
    (c^2 + 1) \bar H_N^2 - (2 \bar H_{\text{w}} + c^2) \bar H_N + \bar H_{\text{w}}^2 = 0,
\end{align}
which can be solved for $\bar H_N$. The two roots of the quadratic equation, which we will denote by $\{h_1,h_2\}$ with $h_1<h_2$, corresponding $\rho_{_{H,R_{\text{w}}}}=r$
and $\rho_{_{H,R_{\text{w}}}}=-r$. Since we know the sign of $r$, there will be no ambiguity on which root to take.

We note that, by setting $z = r \sqrt{N}$ and rearranging (Equation \ref{eq:pre_square}), we have
\begin{align}
    \frac{\He_{\text{w}} - \He_N}{\sqrt{\frac{1-f}{n}\cdot \He_N(1 - \He_N)}} = z,
\end{align}
where $f=n_{\text{w}}/N$.
One may recognize that is the quantity for constructing the classical Wilson score confidence interval for a binomial proportion \cite{brown2001interval}, but with the finite-population correction factor $(1-f)$.  This connection  illuminates the meaning of the particular value of \textit{ddc} ($\rho_{_{H, R_\text{w}}}$) in this context: the quantity $z$, which directly depends on \textit{ddc}, is the corresponding \emph{quantile} used in the Wilson interval. In other words, $z$ is the multiplier or yardstick of the benchmark error (provided by simple random sampling) to measure the error in the estimator $\He_{\text{w}}$.  The fact that it grows with  $\sqrt{N}$, when $\rho_{_{H, R_\text{w}}}$ does not vanish with $1/\sqrt{N}$,  is precisely the explanation  from the \emph{ddc} framework.

\subsection{Scenario estimates}

We focus on three scenarios defined by ranges of $\lambda$ that correspond to three mechanisms:

This allocation scheme allows us to pose scenarios implied by values of $\lambda$ that capture three plausible mechanisms driving bias. First, if hesitant ($H$) and willing ($W$) individuals are equally under-represented ($\lambda \approx 0.5$), leading to over-representation of uptake, correcting for data quality implies that both Willingness and Hesitancy are higher than what surveys report (Extended Data Fig.~\ref{fig:scenario-allsurveys}, yellow bands). We label this the \textit{uptake} scenario because, among the three components, uptake has the largest absolute \emph{ddc}. Alternatively, the under-representation of the \textit{hesitant} population could be the largest source of bias, possibly due to under-representation of people with low institutional trust who may be less likely to respond to surveys and more likely to be hesitant. This implies $\lambda \approx 0$ and is shown in the red bands. The last scenario addresses issues of \textit{access}, where under-representation of people who are willing but not yet vaccinated is the largest source of bias, perhaps due to correlation between barriers to accessing both vaccines and online surveys (e.g., lack of internet access). This implies $\lambda\approx 1$ and upwardly corrects willingness, but does not change hesitancy.

In particular, the values used to generate the bands shown in Extended Data Fig~\ref{fig:scenario-allsurveys} use the following values of lambda:
\begin{itemize}
    \item\textit{Access} (blue bands): $\lambda \in[1,1.2]$, and thus $\rho_{W} \in [-1.2\rho_{V}, -\rho_{V}]$ and $\rho_{H} \in [0, 0.2\rho_{V}]$.
    \item \textit{Hesitancy} (red bands): $\lambda \in [-1.2,-1]$, and thus $\rho_{H} \in [-1.2\rho_{V}, -\rho_{V}]$ and $\rho_{W} \in [0, 0.2\rho_{V}]$. 
    \item \textit{Uptake} (yellow bands): $\lambda \in[0.4,0.6]$ $\rho_{H} \in [-0.6\rho_{V}, -0.4\rho_{V}]$ and $\rho_{W} \in [-0.6\rho_{V}, -0.4\rho_{V}]$.
\end{itemize}

For each of the scenarios we estimates, adjustments with $\rho_{R, V}$ (\emph{ddc} of vaccination) by each survey puts the three survey's estimates of Hesitancy and Willingness in agreement.
Because the width of each band is proportional to each survey's estimated $\rho_{R, V}$ by a constant $\lambda$, it makes sense that Delphi-Facebook has the widest band and Axios-Ipsos has the narrowest band.

The \textit{hesitancy} scenario suggests that the actual rate of hesitancy is about 31-33\% in the most recent waves of Delphi-Facebook and Census Household Pulse, almost double that of original estimates. In the \textit{uptake} scenario, both hesitancy and willingness are about  5 percentage points higher than each survey's original estimates. The \textit{access} scenario suggests that willingness is as high as 21\%, i.e. that a fifth of the US population still faced significant barriers to accessing vaccines as of late May. 

Axios-Ipsos scenarios differ from those of the other two surveys due to its small \textit{ddc}, and different question wording. 
The question that Axios-Ipsos uses to gauge vaccine hesitancy is worded differently from the questions used in Census Household Pulse and Delphi-Facebook. The question asks about likelihood of receiving a ``first generation'' COVID-19 vaccine, which may increase levels of hesitancy among respondents if they believe the survey is asking about an experimental, rather than a thoroughly tested, vaccine. We do see that Axios-Ipsos has markedly higher baseline levels of hesitancy than either Census Household Pulse or Delphi-Facebook. While this is likely driven in part by the lower estimated rates of vaccine uptake, it is also likely due in part to question wording.  Therefore, we exclude Axios-Ipsos from our scenarios of vaccine hesitancy and willingness.

The \textit{ddc} of Axios-Ipsos is small, its estimates of hesitancy are affected less by these scenarios. Furthermore, the implied level of Hesitancy estimates for Axios-Ipsos is higher than that of the other two polls by 5-10 percentage points in the Access scenario.
In fact Axios-Ipsos'  \emph{original} estimates of Hesitancy are higher than the other polls, above and beyond demographic composition differences (Table \ref{tab:outcome-by-demo}).
This is likely to the wording of the inclusion of ``first generation vaccine'' in Axios-Ipsos' vaccine hesitancy question (Methods section \nameref{sec:SM-data-background}). Because such wording differences may confound the interpretation of the scenarios (\emph{ex ante}), we do not present Axios-Ipsos' results in the same figure as the other two surveys in the main text. To be clear, the vaccination is measured in a different question than Hesitancy (Table \ref{table:methodologies}) and does not affect our presentation of vaccination-related outcomes in earlier parts of the article.

This analysis alone cannot determine which scenario is most likely, and scenarios should be validated with other studies. However, we hope that these substantive, mechanism-driven scenarios are useful for policymakers who may need to choose whether to devote scarce resources to the Willing or Hesitant populations.  Extended Data Fig.~\ref{fig:scenario-allsurveys} also shows that when positing these scenarios through a \emph{ddc} framework, the estimates from Delphi-Facebook and Census Household Pulse disagree to a lesser extent than in the reported estimates (Extended Data Fig.~\ref{fig:hesitancy-state-march} and~\ref{fig:hesitancy-state-may}).

\end{document}